\documentclass[%
 reprint,
 amsmath,amssymb,
 aps,
]{revtex4-2}

\usepackage{aasmacros}
\usepackage{natbib}
\usepackage{subcaption}
\usepackage{graphicx}% Include figure files
\usepackage{dcolumn}% Align table columns on decimal point
\usepackage{bm}% bold math
\usepackage{xcolor}
\newcommand{\diff}{\ensuremath{\mathrm{d}}}

\newcommand{\reply}[1]{\textcolor{black}{{#1}}}
\begin{document}

\title{Cosmology with Standard Sirens at Cosmic Noon}

\author{Christine Ye}
\affiliation{Eastlake High School, 400 228th Ave NE, Sammamish, WA 98074, USA}

\author{Maya Fishbach}
\altaffiliation{NASA Hubble Fellowship Program Einstein Postdoctoral Fellow}
\affiliation{Center for Interdisciplinary Exploration and Research in Astrophysics (CIERA) and Department of Physics and Astronomy,
Northwestern University, 1800 Sherman Ave, Evanston, IL 60201, USA}

\date{\today}

\begin{abstract}
  Gravitational waves (GWs) from merging black holes and neutron stars directly measure the luminosity distance to the merger, which, when combined with an independent measurement of the source's redshift, provides a novel probe of cosmology. The proposed next generation of ground-based GW detectors, Einstein Telescope and Cosmic Explorer, will detect tens of thousands of binary neutron stars (BNSs) out to cosmological distances ($z>2$), beyond the peak of the star formation rate (SFR), or ``cosmic noon." At these distances, it will be challenging to measure the sources' redshifts by observing electromagnetic (EM) counterparts or statistically marginalizing over a galaxy catalog. \reply{In the absence of an EM counterpart or galaxy catalog, \citet{2019JCAP...04..033D} showed that theoretical priors on the merger redshift distribution can be used to infer parameters in a $w$CDM cosmology. We argue that in the BNS case, the redshift distribution will be measured by independent observations of short gamma ray bursts (GRBs), kilonovae, and known BNS host galaxies. In particular, the peak redshift will provide a clear feature to compare against the peak distance of the GW source distribution and reveal the underlying redshift-distance relation. We show that, in addition to measuring the background cosmology, this method can constrain the effects of dark energy on modified GW propagation.} As a simple example, we consider the case in which the BNS rate is \textit{a priori} known to follow the SFR. If the SFR is perfectly known, $\mathcal{O}(10,000)$ events (to be expected within a year of observation with Cosmic Explorer) would yield a sub-tenth percent measurement of the combination $H_0^{2.8}\Omega_M$ in a flat $\Lambda$CDM model. \reply{Meanwhile, fixing $H_0$ and $\Omega_M$ to independently-inferred values,} this method may enable a 5\% measurement of the dark energy equation of state parameter $w$ \reply{in a $w$CDM model}. Fixing the background cosmology and instead probing modified GW propagation, the running of the Planck mass parameter $c_M$ may be measured to $\pm0.02$. Although realistically, the redshift evolution of the merger rate will be uncertain, prior knowledge of the peak redshift will provide valuable information for standard siren analyses.
\end{abstract}

\maketitle

\section{Introduction}
\label{sec:intro}
As first pointed out by~\citet{1986Natur.323..310S}, GWs from the coalescence of two compact objects, like black holes and neutron star mergers, provide an absolute distance measurement to the source. In analogy to ``standard candles," like Type Ia supernovae, these GW sources are known as ``standard sirens"~\citep{2005ApJ...629...15H}. If the redshift corresponding to the GW source can also be determined, it is possible to constrain the distance-redshift relation and thereby measure cosmological parameters. However, the redshift cannot be directly extracted from the GWs, because the redshift of the GW signal is degenerate with the mass of the system. Standard siren cosmology therefore relies on external data to infer the redshift of the GW source. 

The most straightforward approach to determine the redshift of a GW observation is to identify an associated EM counterpart, like a short GRB or a kilonova, which in turn allows for a unique host galaxy identification and redshift determination~\citep{2005ApJ...629...15H,PhysRevD.74.063006,2010ApJ...725..496N}. A counterpart standard siren measurement was first carried out following the spectacular multi-messenger detection of the BNS merger GW170817~\citep{2017PhRvL.119p1101A,2017ApJ...848L..12A}, yielding a measurement of the Hubble constant $H_0 = 70^{+12}_{-8}$ km s$^{-1}$ Mpc$^{-1}$~\citep{2017Natur.551...85A}. \citet{2018Natur.562..545C} and \citet{Feeney:2018mkj} showed that 50 detections of BNS mergers by the Advanced LIGO~\citep{2015CQGra..32g4001L} and Virgo~\citep{2015CQGra..32b4001A} GW detector network with associated EM counterparts will enable a $\sim2\%$ measurement of $H_0$, which would provide an important test of the $\Lambda$CDM cosmological model and may help shed light on the persistent $H_0$ tension~\citep{2014A&A...571A..16P,2019ApJ...876...85R,2019ApJ...882...34F,2020PhRvD.101d3533K,2021arXiv210101372B,2021arXiv210301183D}. %At design sensitivity, the current GW detectors will observe BNS mergers out to a few hundred Mpc~\citep{2018LRR....21....3A}, and so the corresponding standard siren measurements will only be sensitive to $H_0$.

Nevertheless, the majority of GW events do not have identified EM counterparts. In the absence of a counterpart, it is possible to statistically marginalize over the redshifts of all of the potential host galaxies in the GW localization volume using a galaxy catalog~\citep{1986Natur.323..310S,PhysRevD.77.043512,PhysRevD.86.043011,2016PhRvD..93h3511O}. This statistical standard siren approach has been applied to several GW events~\citep{2019ApJ...871L..13F,2019ApJ...876L...7S,2019arXiv190806060T,2020ApJ...900L..33P,2021arXiv210112660F}. The most promising dark sirens for the statistical method are nearby, well-localized events, where the number of galaxies in the volume is relatively small and available galaxy catalogs are relatively complete~\citep{2018Natur.562..545C,2019arXiv190806060T,2020PhRvD.101l2001G,2021arXiv210112660F}. When catalogs are incomplete but GW events are well-localized, it may be possible to compare the spatial clustering of GW sources and galaxies as a function of redshift to infer cosmological parameters~\citep{PhysRevD.77.043512,Mukherjee:2018ebj,2020arXiv200501111V,2020ApJ...902...79B,2021PhRvD.103d3520M}. Finally, in the absence of counterparts or catalogs, several authors have proposed GW-only standard siren analyses. Known properties of the source population, such as features in the source-frame mass distribution~\citep{1993ApJ...411L...5C,2012PhRvD..85b3535T,2012PhRvD..86b3502T,2019ApJ...883L..42F,2020arXiv200400036Y,2020arXiv200602211M} or knowledge of the neutron star equation of state~\citep{PhysRevLett.108.091101,2017PhRvD..95d3502D}, can be used to extract the redshift from the observed GW frequency. \citet{2019JCAP...04..033D} pointed out that even if the redshifts of individual GW events cannot be identified, it is possible to extract cosmological information from a population of standard sirens if their redshift distribution is theoretically known from population synthesis simulations. 

In this study, we build on \citet{2019JCAP...04..033D} and further explore the potential of standard siren cosmology without counterparts. We argue that external EM observations, not necessarily associated with GW events, provide a measurement of the BNS redshift distribution that can be leveraged in a standard siren measurement. For example, if the BNS merger rate is known to follow the SFR \reply{with short typical time delays~\citep{2014MNRAS.442.2342D,2016A&A...594A..84G,2019MNRAS.486.2896S}}, we will know that there is an abundance of BNS host galaxies near the peak of the SFR at $z \sim 2$~\citep{2014ARA&A..52..415M,2015MNRAS.447.2575V} without comparing a galaxy catalog against GW events.

This method would be particularly relevant for the next generation of ground-based GW observatories, the proposed detectors Cosmic Explorer~\citep{2015PhRvD..91h2001D} and Einstein Telescope~\citep{2010CQGra..27h4007P,2012CQGra..29l4013S}, which are currently under consideration.
These third-generation (3G) detectors would dramatically increase the distance out to which BNS mergers can be observed, from a few hundred Mpc with current detectors~\citep{2018LRR....21....3A,Chen_2021} to tens of Gpc~\citep{2010CQGra..27u5006S,2019CQGra..36v5002H,2019JCAP...08..015B}.
The 3G detectors will thus most likely observe these mergers past the peak redshift of the merger rate distribution. Depending on the detector network, the BNS rate, and the mass distribution, they will observe on order of $10^5$ BNSs annually~\citep{2019JCAP...08..015B}. \reply{Although some of these GW signals will overlap, the parameters of these sources can nevertheless be measured reliably~\citep{Samajdar:2021egv, pizzati2021bayesian, Himemoto:2021ukb}.}
This large GW dataset will provide a novel probe of the high-redshift universe~\citep{2019BAAS...51c.242K}.
For example, assuming the distance-redshift relation is known, the distribution of their luminosity distances will enable precise measurements of the time delay distribution between star formation and compact object merger~\citep{2012PhRvD..86b3502T,2019ApJ...886L...1V,2019ApJ...878L..13S}. 
Another natural application of 3G detectors is standard siren cosmology out to high redshifts, which can provide independent constraints on dark energy, alternative cosmological models and modified gravity~\citep{2010CQGra..27u5006S,Zhao_2011,2012PhRvD..86b3502T,Cai_2017,2018PhRvD..98b3502N,Zhang_2019,2019JCAP...08..015B,2020arXiv200400036Y,2020JCAP...03..051J,2020arXiv200702883B,2021PhRvD.103d4024P,2021arXiv210301923Y}. However, at $z > 1$, it will become increasingly difficult to observe EM counterparts, both because of their reduced apparent brightness and the large GW localization areas~\citep{2021ApJ...908L...4C}. The statistical method will also face challenges, because galaxy catalogs will be increasingly incomplete at high redshift. GW-only methods drawing on knowledge of the source-frame population, such as the BNS mass distribution~\citep{2012PhRvD..85b3535T,2012PhRvD..86b3502T} or the pair-instability feature in the BBH mass distribution~\citep{2019ApJ...883L..42F,2020arXiv200400036Y} may prove useful; the latter technique may even provide an $\mathcal{O}(10\%)$ measurement of the dark energy equation of state with the current GW detector network~\citep{2019ApJ...883L..42F}. However, these methods rely on some understanding of the evolution of the source population with redshift, which remains observationally and theoretically uncertain~\citep{2021arXiv210107699F}. 

\begin{figure*}
\begin{subfigure}{0.45\textwidth}
\includegraphics[width=\linewidth]{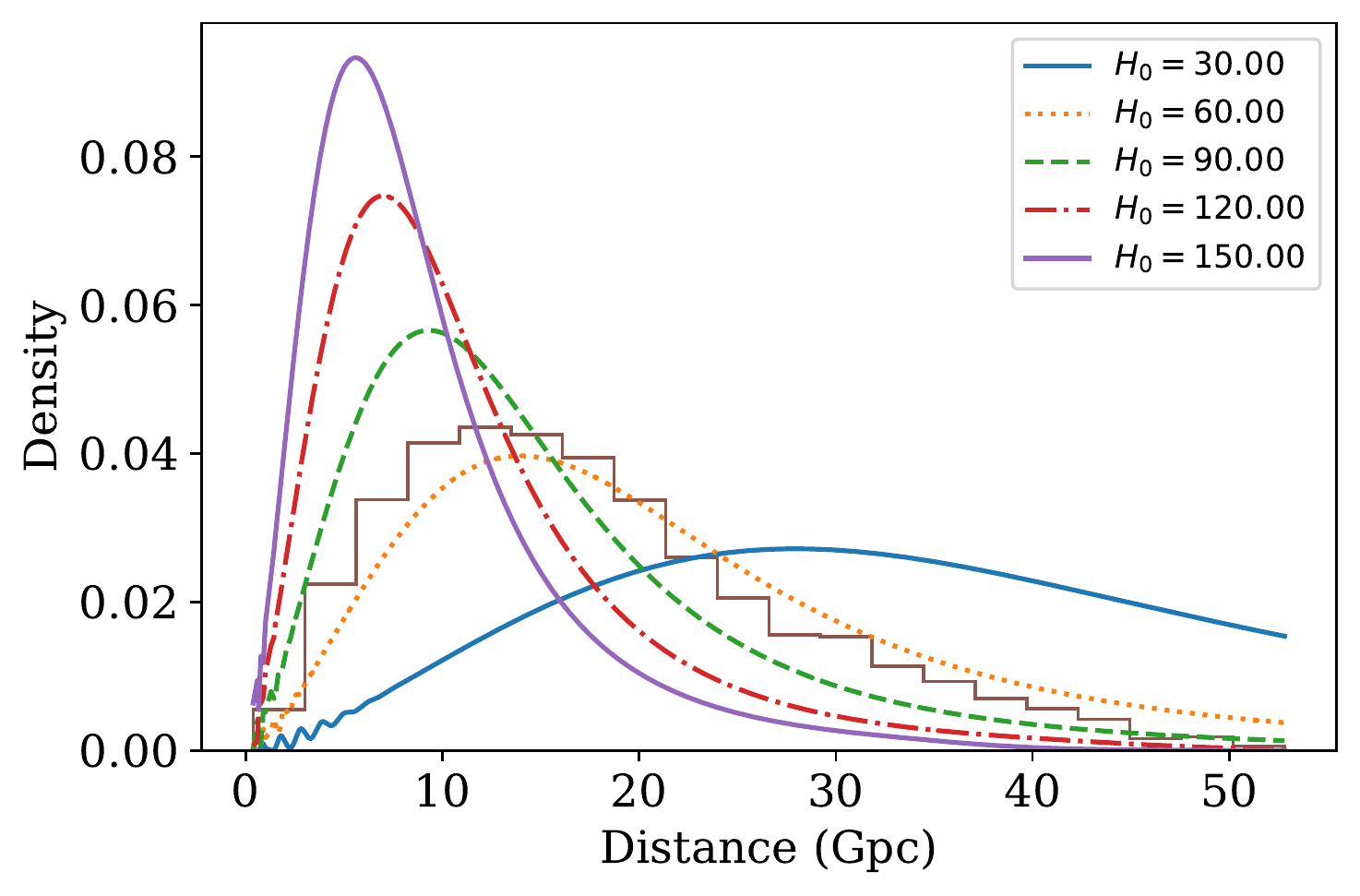}
\caption{Distribution of source luminosity distances for different values of the Hubble constant $H_0$. Smaller values of $H_0$ result in larger observed distances on average.}
\end{subfigure}
\hfill
\begin{subfigure}{0.45\textwidth}
\includegraphics[width=\linewidth]{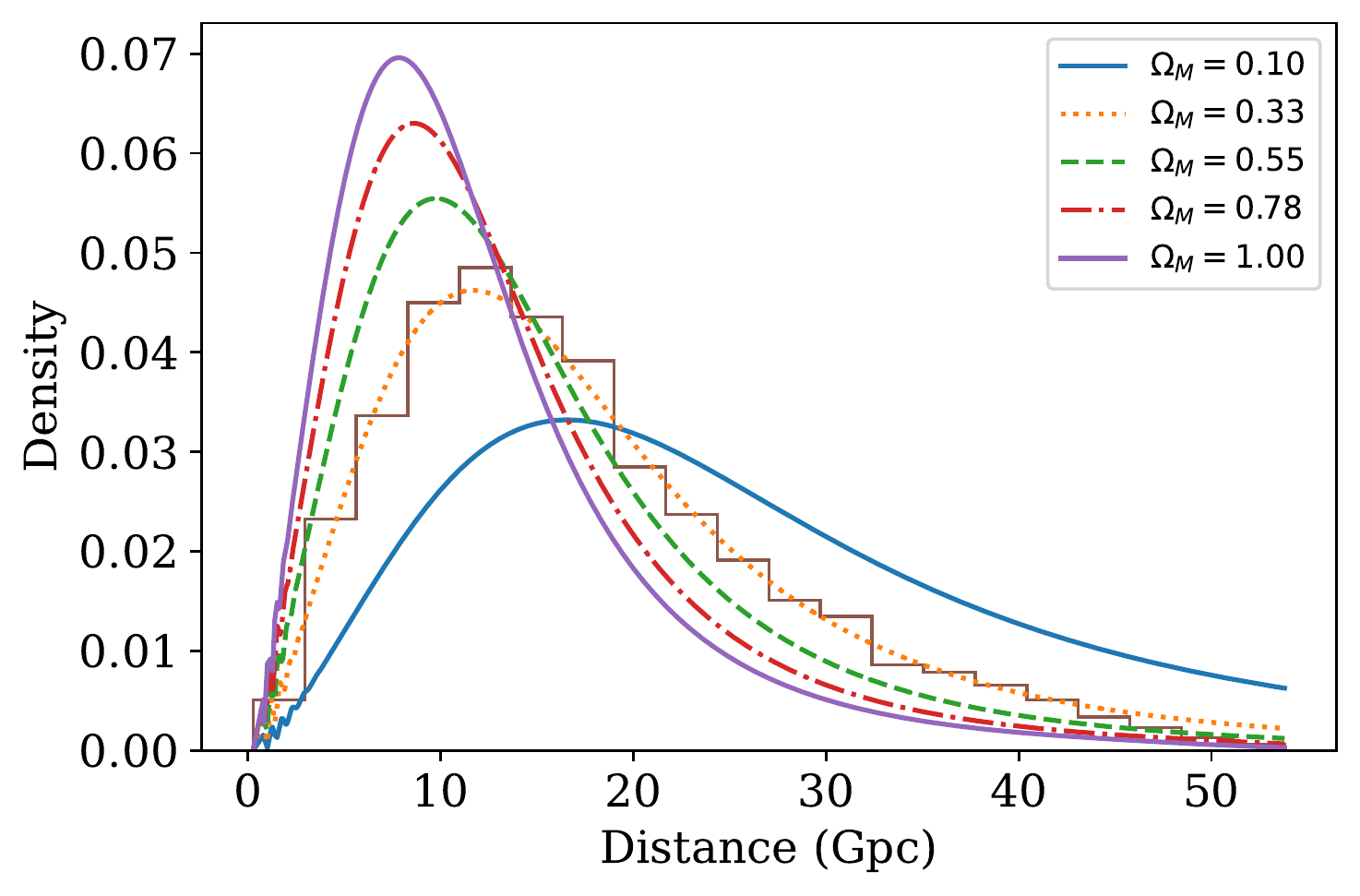}
\caption{Distribution of source luminosity distances for different values of the matter density $\Omega_M$. Smaller values of $\Omega_M$ result in larger observed distances on average.}
\end{subfigure}

\bigskip 
\begin{subfigure}{0.45\textwidth}
\includegraphics[width=\linewidth]{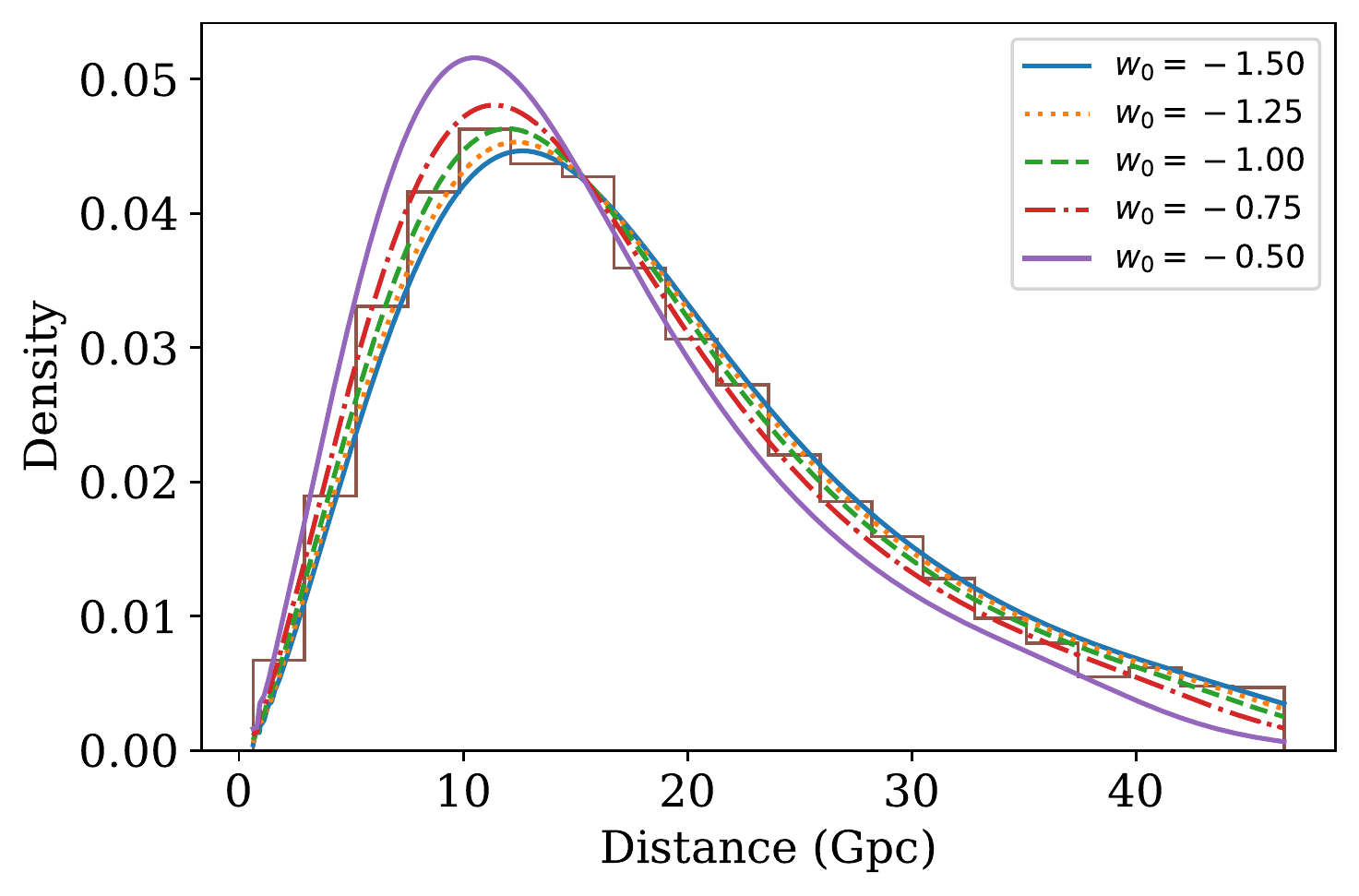}
\caption{Distribution of source luminosity distances for different values of the dark energy equation of state parameter $w_0$ (with $w_a$ fixed to zero). The effect is small compared to the influence of $\Omega_M$ and $H_0$, but visible. Smaller (more negative) values of $w_0$ result in larger observed distances on average.}
\end{subfigure}
\hfill % maximize the horizontal distance between the graphs
\begin{subfigure}{0.45\textwidth}
\includegraphics[width=\linewidth]{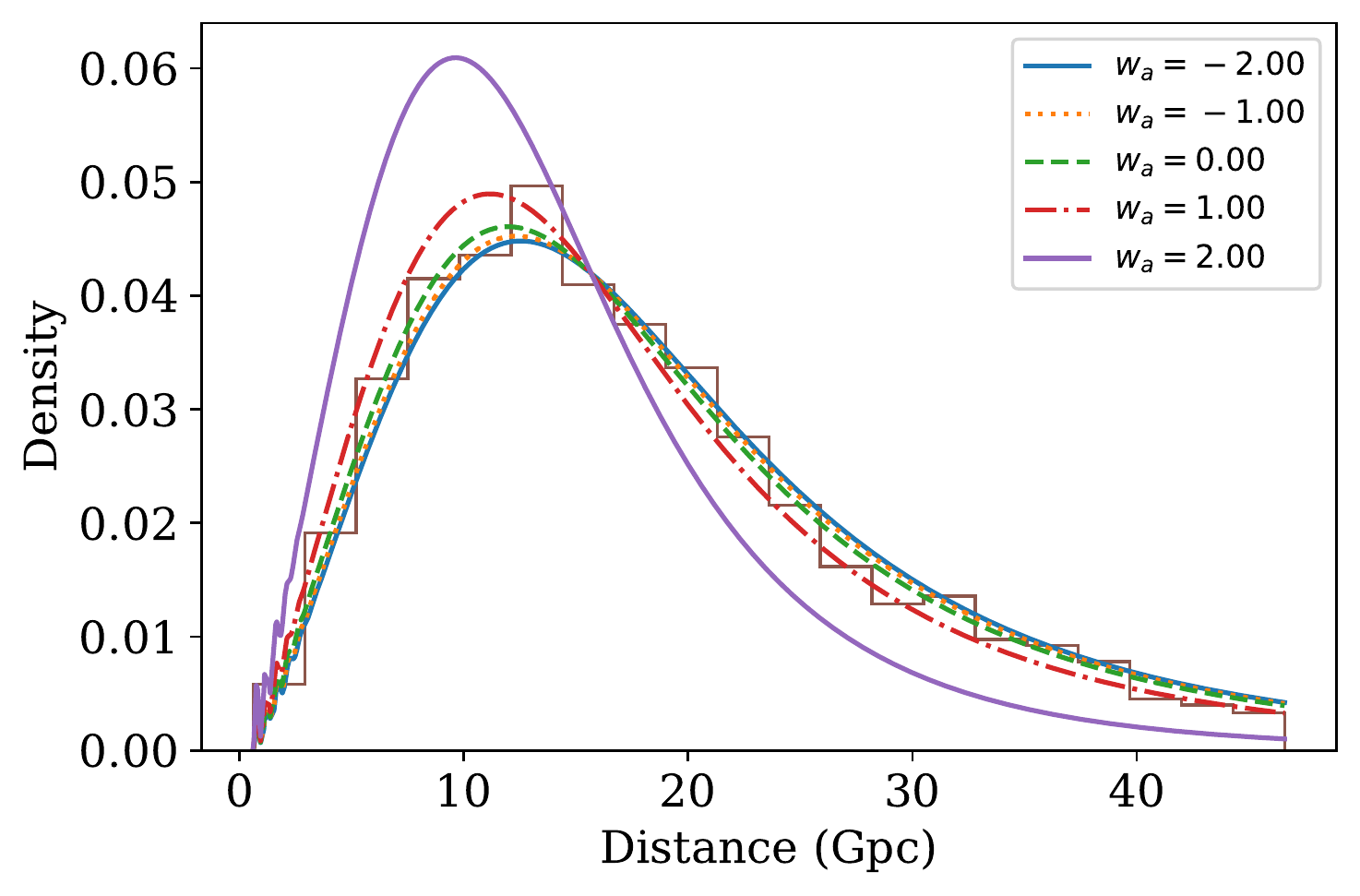}
\caption{Distribution of source luminosity distances for different values of the dark energy equation of state parameter $w_a$ (with $w_0$ fixed to -1). The luminosity distance distribution is much more sensitive to variations towards positive $w_a$ than negative $w_a$. Positive $w_a$ results in smaller observed distances on average.}
\end{subfigure}

\begin{subfigure}{0.45\textwidth}
\includegraphics[width=\linewidth]{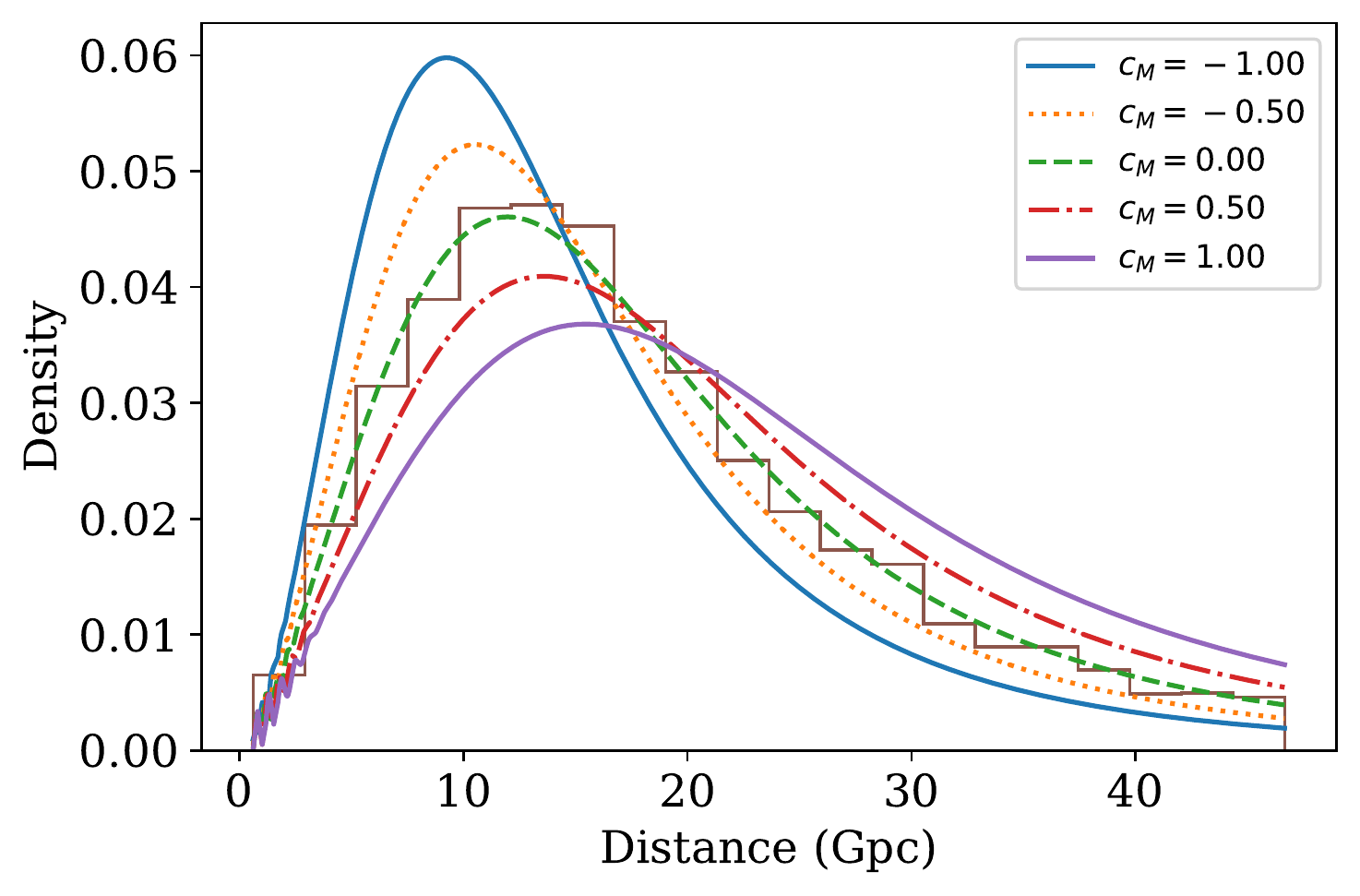}
\caption{Distribution of GW luminosity distances for different values of $c_M$, the parameter controlling the running of the Planck mass in a modified gravity theory, in a flat $\Lambda$CDM background cosmology. Positive $c_M$ results in larger observed distances on average.}
\end{subfigure}

\caption{\label{fig:dLs_cosmo} Effect of cosmological and modified gravity parameters on the GW luminosity distance distribution, for a fixed merger rate evolution $\mathcal{R}(z)$ that follows the Madau-Dickinson SFR. We assume a flat background $w_0w_a$CDM cosmology, with possible modifications to GW propagation parameterized by $c_M$. Each panel varies one parameter at a time, fixing the other parameters to a ``concordance cosmology" of $\Lambda$CDM with $H_0$ = 70 km s$^{-1}$ Mpc$^{-1}$, $\Omega_M = 0.3$, $\Omega_\Lambda = 0.7$. Histograms show simulated distance measurements with concordance cosmology.}
\end{figure*}

These anticipated challenges for standard siren cosmology in the 3G era motivate us to consider supplementary sources of redshift information. Although we cannot always observe the EM counterpart to a GW detection of a BNS, we nevertheless have an external EM sample of short GRBs and kilonovae, the progenitors of which are probably BNS (or neutron-star black hole) mergers\reply{~\citep{2017ApJ...848L..12A}}. This sample will grow in the coming years with the improved sensitivity of upcoming/proposed observing facilities like the Vera Rubin Observatory, JWST and the Roman Space Telescope for observing kilonovae~\citep{2018ApJ...852L...3S} and BurstCube, SVOM, THESEUS and ULTRASAT for observing GRBs and afterglows, among many other proposed missions. We expect that external EM observations of short GRBs and kilonovae will constrain the number density of these sources as a function of redshift\reply{, either by accumulating a large sample of sources with known redshifts, as identified through afterglow or host galaxy spectroscopy, or jointly fitting the GRB observed flux distribution to the underlying luminosity function and redshift distribution  ~\citep{2011ApJ...727..109V,2014MNRAS.442.2342D,2015MNRAS.448.3026W,2016A&A...594A..84G,2018MNRAS.477.4275P,2021arXiv210413555D}}. Even observations within a limited redshift range can provide valuable information about the redshift evolution of the merger rate if host galaxies can be identified~\citep{2013ApJ...776...18F}. The properties of host galaxies (e.g. their masses and specific star formation rates) can be used to measure the time delay distribution~\citep{2010ApJ...725.1202L,2019ApJ...878L..12S,2020ApJ...905...21A} and therefore, assuming some knowledge about the cosmic SFR, the BNS merger rate as a function of redshift. This measurement may be assisted by theoretical guidance that the BNS progenitor formation rate probably traces the SFR, independent of the (largely uncertain) metallicity evolution~\citep{2020ApJ...898..152S}. \reply{Additional information about the BNS merger rate evolution can be gained by studying the Galactic evolution of r-process elements~\citep{2019MNRAS.486.2896S}.} 

Figure~\ref{fig:dLs_cosmo} illustrates how, for a fixed merger rate evolution $\mathcal{R}(z)$, the luminosity distance distribution, as measured with GW standard sirens, depends on cosmology. For simplicity, we fix the underlying merger rate to the Madau-Dickinson SFR (see Eq.~\ref{eq:MD}).
Building upon this intuition, in the following we explore how comparing GW observations of luminosity distances to a known redshift distribution can be used to extract cosmological and modified gravity parameters, including the Hubble constant, the matter density and the dark energy equation of state in a flat $w_0w_a$-CDM cosmology~\citep{2001IJMPD..10..213C,2003PhRvL..90i1301L}, and the running of the Planck mass following the framework of~\citet{2019PhRvD..99h3504L}. Throughout, we fix the geometry of the Universe to be spatially flat, motivated by tight constraints on the curvature from cosmic microwave background (CMB) and baryon acoustic oscillation (BAO) data~\citep{2020MNRAS.496L..91E}.

The general idea is that the GW luminosity distance $D_L^\mathrm{GW}$, as a function of the redshift $z$, depends on these constants that we wish to measure. In general relativity (GR), the luminosity distance $D_L^\mathrm{GW}$ that is extracted from the GW signal is the same luminosity distance $D_L^\mathrm{EM}$ that governs electromagnetic sources, and we have~\citep{1999astro.ph..5116H}:
\begin{equation}
\label{eq:dL1}
    D_L^\mathrm{EM}(z) = (1 + z) \frac{c}{H_0} \int_0^z \frac{dz'}{E(z')},
\end{equation}
where $c$ is the speed of light, $H_0$ is the Hubble constant, and assuming a flat universe, $E(z)$ is:
\begin{equation}
\label{eq:Ez}
    E(z) = \sqrt{\Omega_M (1 + z)^3 + (1 - \Omega_M)I(z) },
\end{equation}
with $\Omega_M$ being the dimensionless matter density today, $(1 - \Omega_M)$ the dark energy density (in a flat universe with a negligible radiation density today), and $I(z)$, in the $w_0w_a$ (CPL model) for the dark energy equation of state, given by~\citep{2001IJMPD..10..213C,2003PhRvL..90i1301L,2018AJ....156..123A}:
\begin{equation}
\label{eq:Iz}
    I(z) = (1 + z)^{3(1 + w_0 + w_a)} \exp\left(-3w_a \frac{z}{1 + z} \right).
\end{equation}
The above reduces to $w$CDM for $w_a = 0$ and $\Lambda$CDM for $w = -1$, $w_a = 0$. We use \textsc{astropy}~\citep{2018AJ....156..123A} for cosmological calculations.

Modified gravity theories~\citep{2012PhR...513....1C,2015PhR...568....1J,2016RPPh...79d6902K}, including models of dynamical dark energy, may alter the amplitude of the GW signal compared to GR in addition to altering the background cosmology away from $\Lambda$CDM, so that the measured $D_L^\mathrm{GW}$ differs from the electromagnetic luminosity distance~\citep{2007ApJ...668L.143D,2015PhLB..742..353P,2016JCAP...03..031L,2018PhRvD..97j4066B,2018JCAP...07..048P,2018JCAP...06..029A,2018JCAP...03..005L,2018FrASS...5...44E,2018PhRvD..97j4066B,2018PhRvD..98b3510B,2019PhRvL.123a1102A,Mukherjee:2019wcg,2019PhRvD..99h3504L,2019PhRvD..99j4038N,Mukherjee:2019wfw,2020PhRvD.102d4009M,2021MNRAS.502.1136M,2021JCAP...02..043M, 2021JCAP...01..068B}. The effect of the GR deviations on GW propagation may be much more significant, and therefore easily measurable with GW events, than the modifications to the background expansion~\citep{2020JCAP...04..010B}. While the multimessenger detection of GW170817 has put tight constraints on the speed of GW propagation, deviations affecting the GW amplitude remain relatively poorly constrained~\citep{2017PhRvL.119y1304E}. In this paper, we consider the example of GW damping caused by an effective running of the Planck mass. Following~\citet{2019PhRvD..99h3504L}, we model the time evolution of the Planck mass with an additional parameter $c_M$ on top of the background cosmology, assumed to follow flat $\Lambda$CDM. The GW luminosity distance $D_L^\mathrm{GW}$ is then the product of Eq.~\ref{eq:dL1} (with $w = -1$, $w_a = 0$ for $\Lambda$CDM) with the extra factor:
\begin{equation}
\label{eq:running}
    \frac{D_L^\mathrm{GW}}{D_L^\mathrm{EM}} = \exp\left(\frac{c_M}{2(1 - \Omega_M)} \ln \frac{1 + z}{\left( \Omega_M ( 1 + z)^3 + 1 - \Omega_M \right) ^{1/3}} \right),
\end{equation}
where $c_M = 0$ reduces to GR, i.e. $D_L^\mathrm{GW} = D_L^\mathrm{EM}$.

The remainder of the paper is organized as follows. Section~\ref{sec:methods} describes the statistical framework that we apply to simulated GW data. We show the results of the simulations in terms of projected constraints in the cosmological parameters in Section~\ref{sec:results}. We conclude in Section~\ref{sec:conclusion}.

\section{Methods}
\label{sec:methods}
This section describes the analysis and simulation methods. We derive the hierarchical Bayesian likelihood for the joint inference of the cosmological parameters and the redshift distribution parameters in Section~\ref{sec:stats} and describe the application of this likelihood to simulated data in Section~\ref{sec:sim}.
\subsection{Statistical framework}
\label{sec:stats}
We assume that the underlying redshift distribution of sources can be described by some parameters $\lambda$ with some additional possible dependence on the cosmological parameters $\mathcal{H}$. We write this as $p(z \mid \lambda, \mathcal{H})$. As a probability density function, $p(z \mid \lambda, \mathcal{H})$ integrates to unity over $0 < z < z_\mathrm{max}$. The population-level parameters are therefore $\lambda$ and $\mathcal{H}$. Often the redshift distribution is expressed as a merger rate density $\mathcal{R}(z)$, which refers to the number of mergers per comoving volume and source-frame time, and can be equivalently written as $\frac{\diff N}{\diff V_c \diff t_s}$ where $V_c$ is the comoving volume and $t_s$ is the source-frame time. The redshift distribution $p(z)$ is related to the redshift-dependent merger rate density $\mathcal{R}(z)$ by:
\begin{equation}
    p(z) \propto \mathcal{R}(z) \frac{\diff V_c}{\diff z} \frac{1}{1 + z}.
\end{equation}
We note that the conversion between $\mathcal{R}(z)$ and $p(z)$ depends on the differential comoving volume element $\frac{\diff V_c}{\diff z}$, which depends on cosmology. Assuming a flat universe,~\citep{1999astro.ph..5116H}:
\begin{equation}
    \frac{dV_c}{dz} = \frac{c}{H_0}\frac{D_L^\mathrm{EM}(z)^2}{(1 + z)^2E(z)},
\end{equation}
with $D_L^\mathrm{EM}(z)$ given by Eq.~\ref{eq:dL1} and $E(z)$ given by Eqs.~\ref{eq:Ez}-\ref{eq:Iz}.
Depending on the type of observations, the measurement of $p(z)$ and/or $\mathcal{R}(z)$ may depend on the assumed cosmology. If we have a redshift catalog of sources; i.e., the number of sources per redshift, we have a direct measurement of $p(z \mid \lambda)$ independent of cosmology. However, if we use observed fluxes to reconstruct the redshift evolution, we may measure $\mathcal{R}(z)$ more directly. The method described below applies to either scenario, but in our simulations we consider the case where a measurement of $\mathcal{R}(z)$ is available.
 
We use a hierarchical Bayesian framework~\citep{2004AIPC..735..195L,2010PhRvD..81h4029M,2019MNRAS.486.1086M} to write the likelihood of the data $d_i$ from event $i$, given the population-level parameters, as:
\begin{align}
\label{eq:single-likelihood}
    p(d_i \mid \lambda, \mathcal{H}) &= \int_0^{z_\mathrm{max}} p(d_i, z_i \mid \lambda, \mathcal{H}) \diff z_i \nonumber \\
    &= \int_0^{z_\mathrm{max}}  p(d_i \mid D_L(z_i, \mathcal{H})) p(z_i \mid \lambda, \mathcal{H}) \diff z_i,
\end{align}
where $D_L(z_i, \mathcal{H})$ denotes the luminosity distance corresponding to the redshift $z_i$ and the cosmology $\mathcal{H}$.
For simplicity of notation, we use $D_L$ to denote the GW luminosity distance $D_L^\mathrm{GW}$ throughout, even when we consider modifications to GR (e.g. Eq.~\ref{eq:running}).
In the above we have implicitly marginalized over any other parameters of the GW signal, so that the marginal likelihood of $d_i$ depends only on the GW luminosity distance $D_L(z_i, \mathcal{H})$. In reality, the GW data also depends on the detector-frame (redshifted) masses of the source; this is discussed further below.

In the presence of GW selection effects, we must modify the likelihood of Eq.~\ref{eq:single-likelihood} to account for the fact that some mergers do not produce detectable data $d_i$. If only data passing some threshold $d^\mathrm{thresh}$ are detected, the likelihood from each event must be normalized by a factor $\beta(\lambda, \mathcal{H})$~\citep{2018Natur.562..545C,2019MNRAS.486.1086M}:
\begin{align}
\label{eq:beta-general}
    \beta(\lambda, \mathcal{H}) &= \\ \nonumber \int_{d > d^\mathrm{thresh}} &\int_0^{z_\mathrm{max}} p(d \mid D_L(z, \mathcal{H}) ) p(z \mid \lambda, \mathcal{H}) \,\diff z \,\diff d.
\end{align}
The single-event likelihood, corrected for selection effects, is then:
\begin{equation}
\label{eq:single-likelihood-selection}
    p(d_i \mid \lambda, \mathcal{H}) = \frac{\int_0^{z_\mathrm{max}}  p(d_i \mid D_L(z_i, \mathcal{H})) p(z_i \mid \lambda, \mathcal{H}) \diff z_i}{\int_{d > d^\mathrm{thresh}} \int_0^{z_\mathrm{max}} p(d \mid D_L(z, \mathcal{H}) ) p(z \mid \lambda, \mathcal{H}) \,\diff z \,\diff d}.
\end{equation}
This differs from the likelihood used in \citet{2019JCAP...04..033D}, which incorporated selection effects by replacing the astrophysical redshift distribution $p(z \mid \lambda, \mathcal{H})$ with the redshift distribution of detected GW events; see \citet{2019MNRAS.486.1086M} for a derivation of the hierarchical Bayesian likelihood in the presence of selection effects.

The total likelihood of $N$ GW events with data $\mathbf{d}$ is the product of the individual-event likelihoods of Eq.~\ref{eq:single-likelihood-selection}:
\begin{equation}
\label{eq:total-likelihood}
    p(\mathbf{d} \mid \lambda, \mathcal{H}) = \prod_{i = 1}^{N} p(d_i \mid \lambda, \mathcal{H} ).
\end{equation}
Using Bayes' rule, we get the posterior on the cosmological parameters $\mathcal{H}$, given some prior $p_0(\mathcal{H})$:
\begin{equation}
\label{eq:posterior}
    p(\mathcal{H} \mid \mathbf{d}, \lambda) \propto p(\mathbf{d} \mid \lambda, \mathcal{H}) p_0(\mathcal{H}).
\end{equation}

In the above, we have made the simplifying assumption that the data (and their detectability) depend on the source's redshift only through the GW luminosity distance.  This is a simplification because in reality, the amplitude and frequency of a signal also depends on the source's redshifted masses and spins; in fact, if we have prior knowledge about the source-frame mass distribution, observing the redshifted masses can by itself probe the distance-redshift relationship~\citep{2012PhRvD..85b3535T,2012PhRvD..86b3502T}. Nevertheless, \reply{because we wish to isolate the information available from the luminosity distance distribution alone,} for this proof-of-principle study we approximate that the GW data depends only on the observed luminosity distance. The masses $m_1(1+z)$ and $m_2(1+z)$ can be easily added into the likelihood of Eq.~\ref{eq:single-likelihood-selection} by considering the GW likelihood $p\left(d \mid D_L(z, \mathcal{H}), m_1(1+z), m_2(1+z)\right)$ and a population model $p(m_1, m_2, z \mid \lambda)$.

We have also ignored the additional distance uncertainty due to the effects of weak gravitational lensing, which will contribute an additional $1\sigma$ uncertainty of $\sim0.05z$ to the measured distance depending on the source redshift $z$. If the distribution of lensing magnifications is known, this contribution can be marginalized over in the GW likelihood without affecting the rest of our formalism~\citep{Holz_2005,2005ApJ...629...15H,PhysRevD.81.124046,2010CQGra..27u5006S,Zhao_2011}. The statistical uncertainties we assume for mock data in the following subsection are large enough to encompass this additional contribution. Alternatively, one can simultaneously fit for the magnification distribution or power spectrum as a function of redshift, which may provide useful constraints on large-scale structure~\citep{PhysRevLett.110.151103,Mukherjee:2019wfw,2019PhRvD..99h3526C}. An additional source of uncertainty will be the calibration uncertainty due in the detector response. This will likely contribute a systematic uncertainty that will limit the accuracy of any standard siren cosmological analyses. 

\subsection{Simulations}
\label{sec:sim}
We apply the likelihood analysis described in the previous subsection~\ref{sec:stats} to mock data.
For simplicity, we assume that the evolution of the merger rate is perfectly known to follow the Madau-Dickinson SFR~\citep{2014ARA&A..52..415M}, peaking at $z \sim 2$:
\begin{equation}
\label{eq:MD}
    \mathcal{R}(z) \propto \frac{(1 + z)^{2.7}}{1 + (\frac{1+z}{2.9})^{5.6}},
\end{equation}
and so the redshift distribution follows:
\begin{equation}
\label{eq:pz-md}
    p(z \mid \lambda, \mathcal{H}) = A \frac{\diff V_c}{\diff z} \frac{1}{1+z}\frac{(1 + z)^{2.7}}{1 + (\frac{1+z}{2.9})^{5.6}} ,
\end{equation}
where $A$ is a normalization constant ensuring that the redshift distribution integrates to unity over the range $0 < z < z_\mathrm{max}$. We take $z_\mathrm{max} = 8$, which ensures that it is larger than the maximum detected BNS distance for any choice of cosmological parameters in our prior. If the maximum astrophysical merger redshift is within the GW detector horizon, it may serve as another feature that can be leveraged for cosmological analyses. We stress that in reality, we do not expect the redshift distribution to be known perfectly, so that instead of using a $\delta$-function prior on $\lambda$ as we effectively assume here, \reply{future measurements} will use a posterior probability distribution on $\lambda$ inferred from external observations.

\begin{figure}
    \centering
    \includegraphics[width=0.5\textwidth,keepaspectratio]{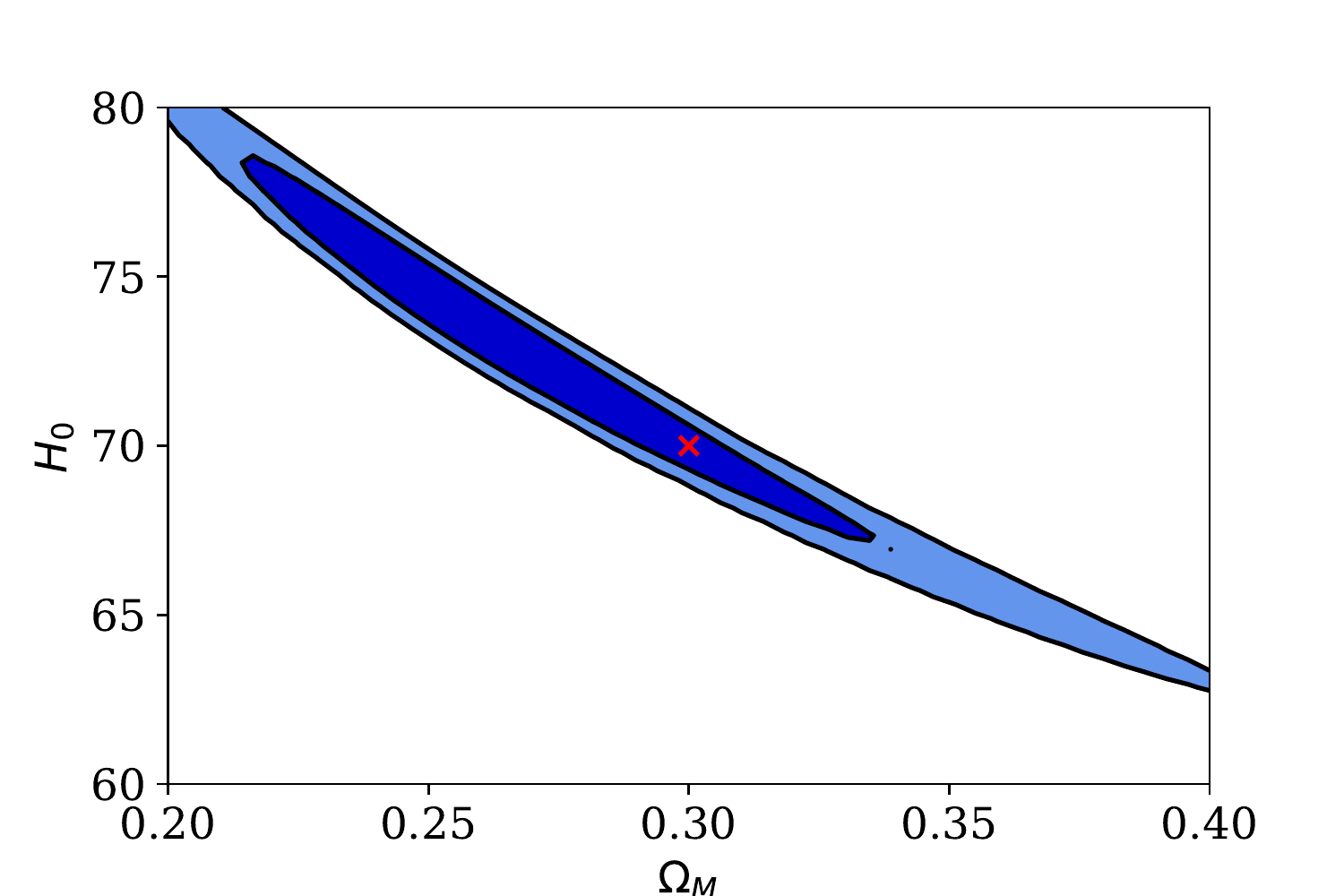}
    \caption{2D posterior distribution in $H_0$ and $\Omega_M$ for a flat $\Lambda$CDM model, inferred from 10,000 simulated luminosity distance measurements with $D_L^\mathrm{max} = 40$ Gpc. We assume flat priors on $H_0$ and $\Omega_M$. 68\% and 95\% contours are shown. Input cosmology is marked with an X.}
    \label{fig:contour}
\end{figure}

\begin{figure*}
        \centering

        \begin{subfigure}[b]{0.3\textwidth}  
            \centering 
            \includegraphics[width=\textwidth]{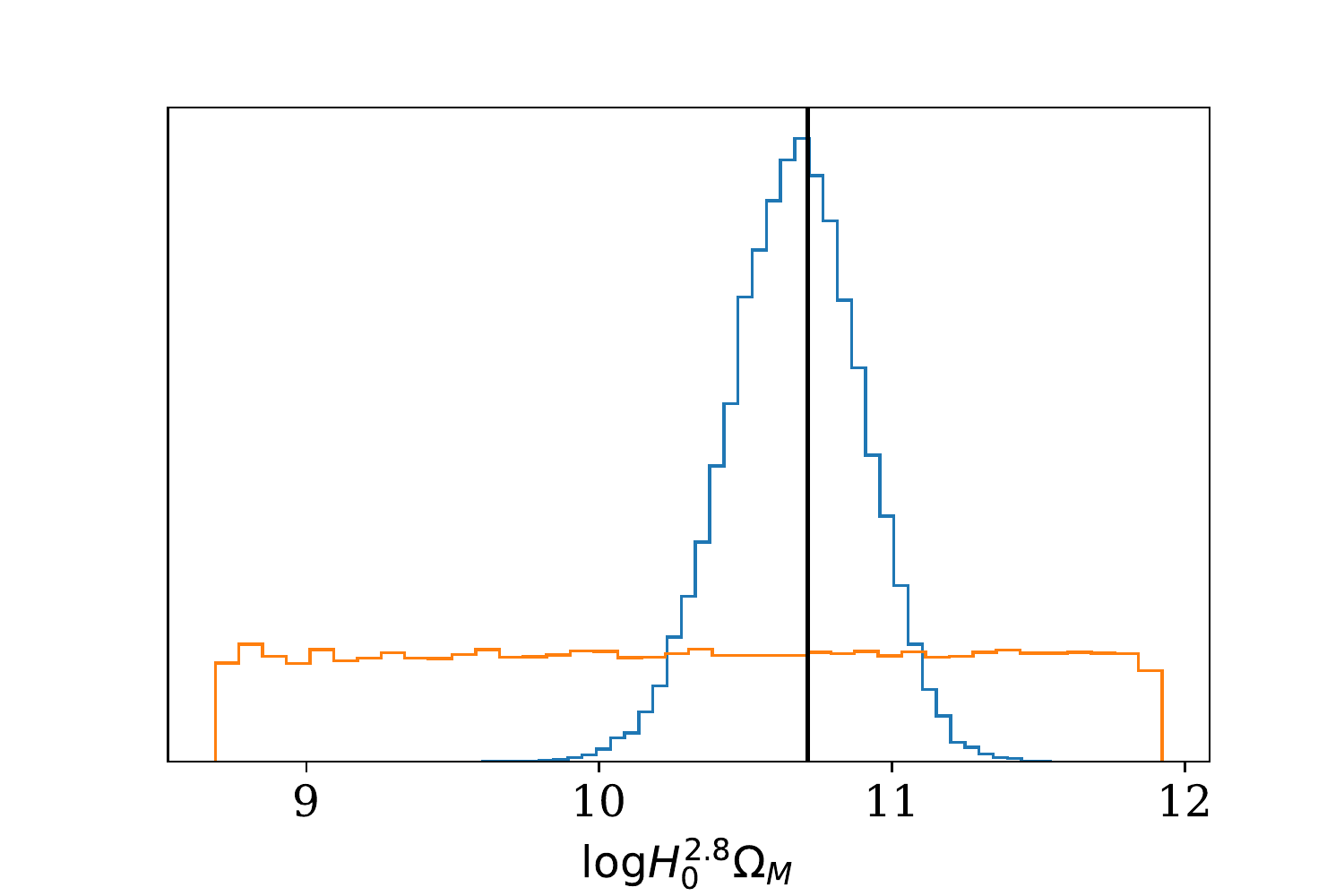}
            \caption[]%
            {{\small 100 measurements}}    
            \label{fig:100}
        \end{subfigure}
        %\vskip\baselineskip
        \begin{subfigure}[b]{0.3\textwidth}   
            \centering 
            \includegraphics[width=\textwidth]{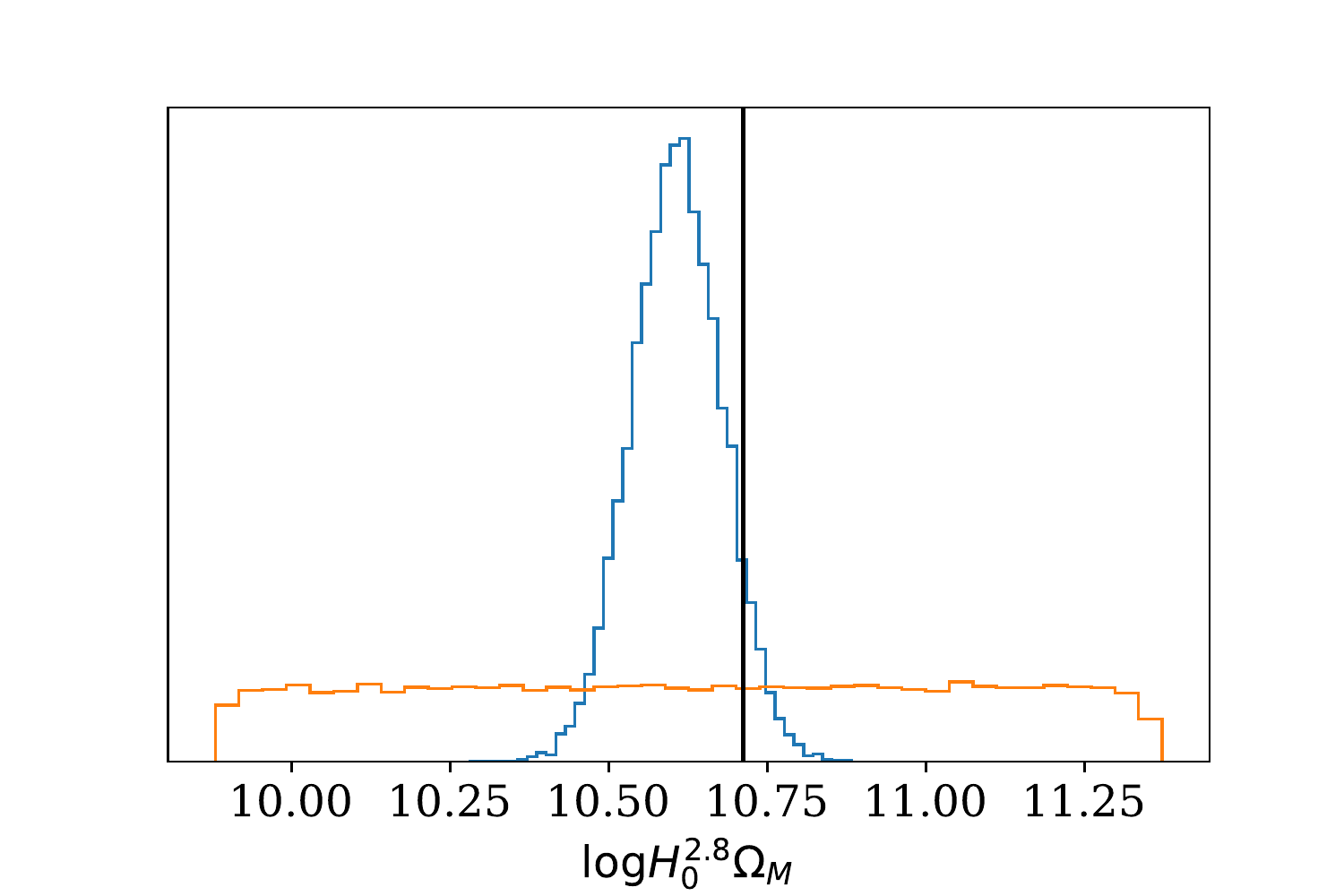}
            \caption[]%
            {{\small 1000 measurements}}    
            \label{fig:1000}
        \end{subfigure}
        %\hfill
        \begin{subfigure}[b]{0.3\textwidth}   
            \centering 
            \includegraphics[width=\textwidth]{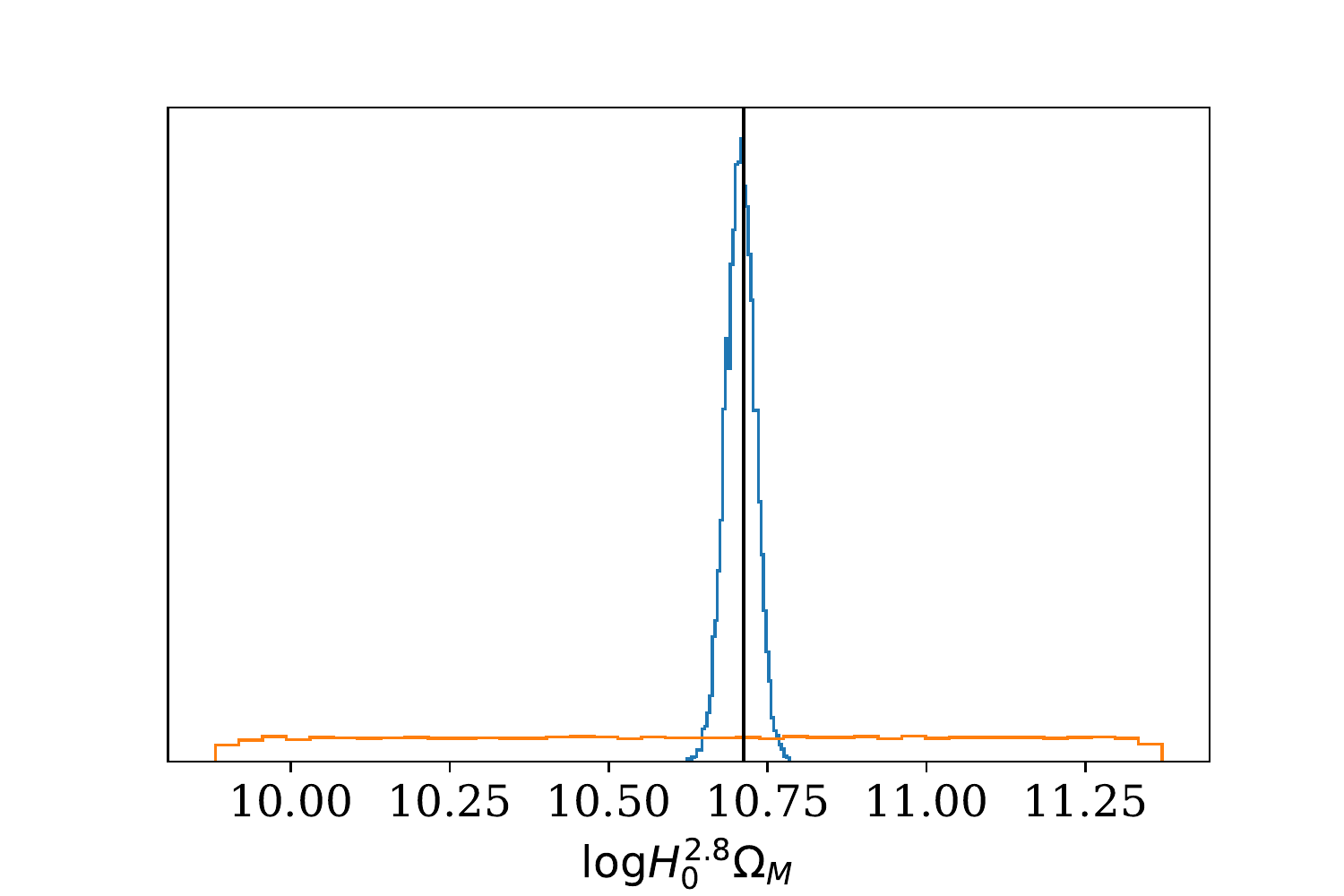}
            \caption[]%
            {{\small 10000 measurements}}    
            \label{fig:10000}
        \end{subfigure}
    \caption[ ]
    { Posterior (blue) and prior (orange) on $H_{0}^{2.81}\Omega_{M}$ inferred from simulated luminosity distances. We reweight the flat prior on $\Omega_M$ and $H_0$ of Fig.~\ref{fig:contour} to a flat prior on the combination $H_{0}^{2.81}\Omega_{M}$.} 
    \label{fig:histograms}
\end{figure*}

For our simulations, we fix a cosmology and draw redshifts $z_i$ from the redshift distribution of Eq.~\ref{eq:pz-md}. Under the fixed cosmology, this gives us the true luminosity distances $D_L^i$ of our mock sources. We then simulate measurement uncertainty, drawing observed distances $D_L^{\mathrm{obs}, i}$ assuming that the GW distance likelihood follows a lognormal distribution with roughly 10\% measurement uncertainty~\reply{(see Appendix B of \citet{2019ApJ...878L..13S})}:
\begin{equation}
\label{eq:DLobs}
    \log D_L^{\mathrm{obs}, i} \sim \mathcal{N}(\mu = \log D_L^i, \sigma = 0.1),
\end{equation}
where $\mathcal{N}(\mu, \sigma)$ denotes the normal distribution with mean $\mu$ and standard deviation $\sigma$.
In other words, we write the GW likelihood $p(d_i \mid D_L(z , \mathcal{H}))$ of Eq.~\ref{eq:single-likelihood-selection} as:
\begin{align}
    &p(d_i \mid D_L(z , \mathcal{H})) = p(D_L^{\mathrm{obs}, i} \mid D_L(z , \mathcal{H})) \\ &\propto \frac{1}{D_L^{\mathrm{obs}, i}} \exp \left(-\frac{1}{2}\left(\frac{\log D_L^{\mathrm{obs}, i} - \log D_L(z, \mathcal{H})}{0.1}\right)^2\right).
\end{align}
This is a conservative assumption compared to parameter estimation simulations and Fisher matrix analyses~\citep{2019JCAP...08..015B,2019ApJ...878L..13S}.
Next we apply selection effects. 
We neglect the effects of the sky-dependent GW detector sensitivity and detector-frame mass (see the discussion in the previous subsection), and simply assume that GW sources are detected if and only if their observed distance is within some maximum $D_L^\mathrm{max}$. We throw out all simulated $D_L^{\mathrm{obs}, i} > D_L^\mathrm{max}$ as below the detection threshold. \reply{As the observed luminosity distance includes a log-normal error term, the detection probability as a function of the true luminosity distance follows a smooth sigmoid function. The detectability of BNS mergers as a function of distance for 3G observatories has large uncertainties, stemming from the BNS mass distribution and details about the 3G detector network. We bound this uncertainty by exploring two choices for the $D_L^\mathrm{max}$ parameter, 20 Gpc and 40 Gpc.  These roughly correspond to Cosmic Explorer's 50\% ``response distance," or the distance at which 50\% of sources are detectable~\citep{Chen_2021}, for binaries with total source-frame masses of $3\,M_\odot$ and $4\,M_\odot$, respectively (see Fig. 1 of \citet{2019CQGra..36v5002H}, assuming a \textit{Planck} 2018 cosmology).}

Again writing $p(d \mid D_L(z, \mathcal{H}) ) = p(D_L^\mathrm{obs} \mid D_L(z, \mathcal{H}) )$, Eq.~\ref{eq:beta-general} then becomes:
\begin{equation}
    \beta(\lambda, \mathcal{H}) = \int_0^{D_L^\mathrm{max}} \int_0^{z_\mathrm{max}}  p(D_L^\mathrm{obs} \mid D_L(z, \mathcal{H}) ) p(z \mid \lambda) \, \diff z \, \diff D_L^\mathrm{obs}.
\end{equation}
Under the assumption that $p(D_L^\mathrm{obs} \mid D_L )$ is a log-normal distribution, we can simplify the integral over $D_L^\mathrm{obs}$:
\begin{align}
\label{eq:beta-specific}
    &\beta(\lambda, \mathcal{H}) = \nonumber \\ 
    &\int_0^{z_\mathrm{max}}  \frac{1}{2} \left( 1 + erf \left(\frac{\log D_L^\mathrm{max} - \log D_L(z, \mathcal{H})}{\sqrt{2}\sigma}\right) \right) p(z \mid \lambda) \diff z,
\end{align}
where $erf(x)$ is the error function and we have picked $\sigma = 0.1$.

For all the $D_L^\mathrm{obs, i}$ that are ``detected," we compute the likelihood of Eq.~\ref{eq:single-likelihood-selection}. The final posterior probability on the cosmological parameters $\mathcal{H}$ is proportional to the product of these likelihoods multiplied by the prior on $\mathcal{H}$, as in Eq.~\ref{eq:posterior}.

\section{Results}
\label{sec:results}

\begin{figure*}
        \centering
        \begin{subfigure}[b]{0.475\textwidth}
            \centering
            \includegraphics[width=\textwidth]{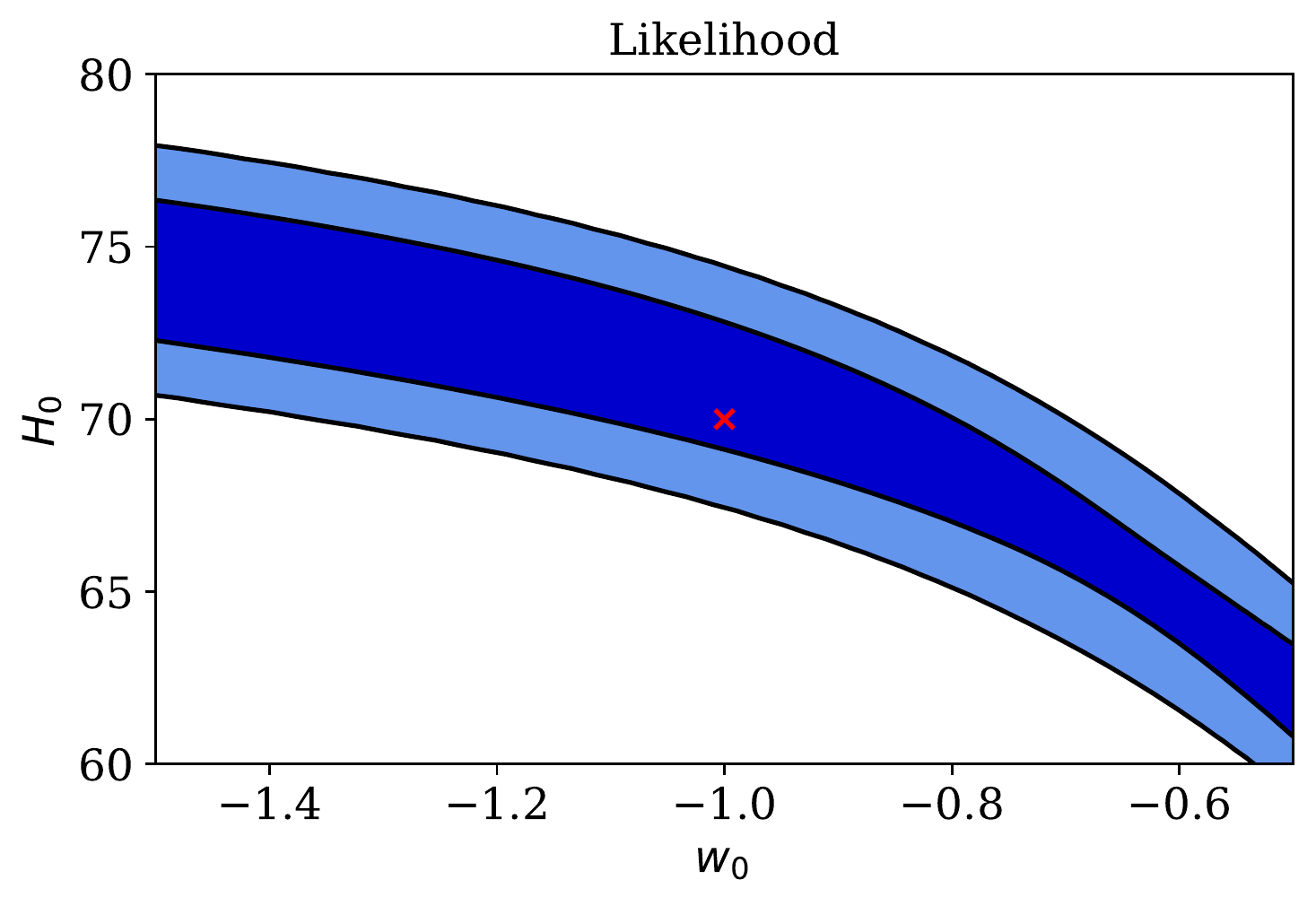}
        \end{subfigure}
        \hfill
        \begin{subfigure}[b]{0.475\textwidth}  
            \centering 
            \includegraphics[width=\textwidth]{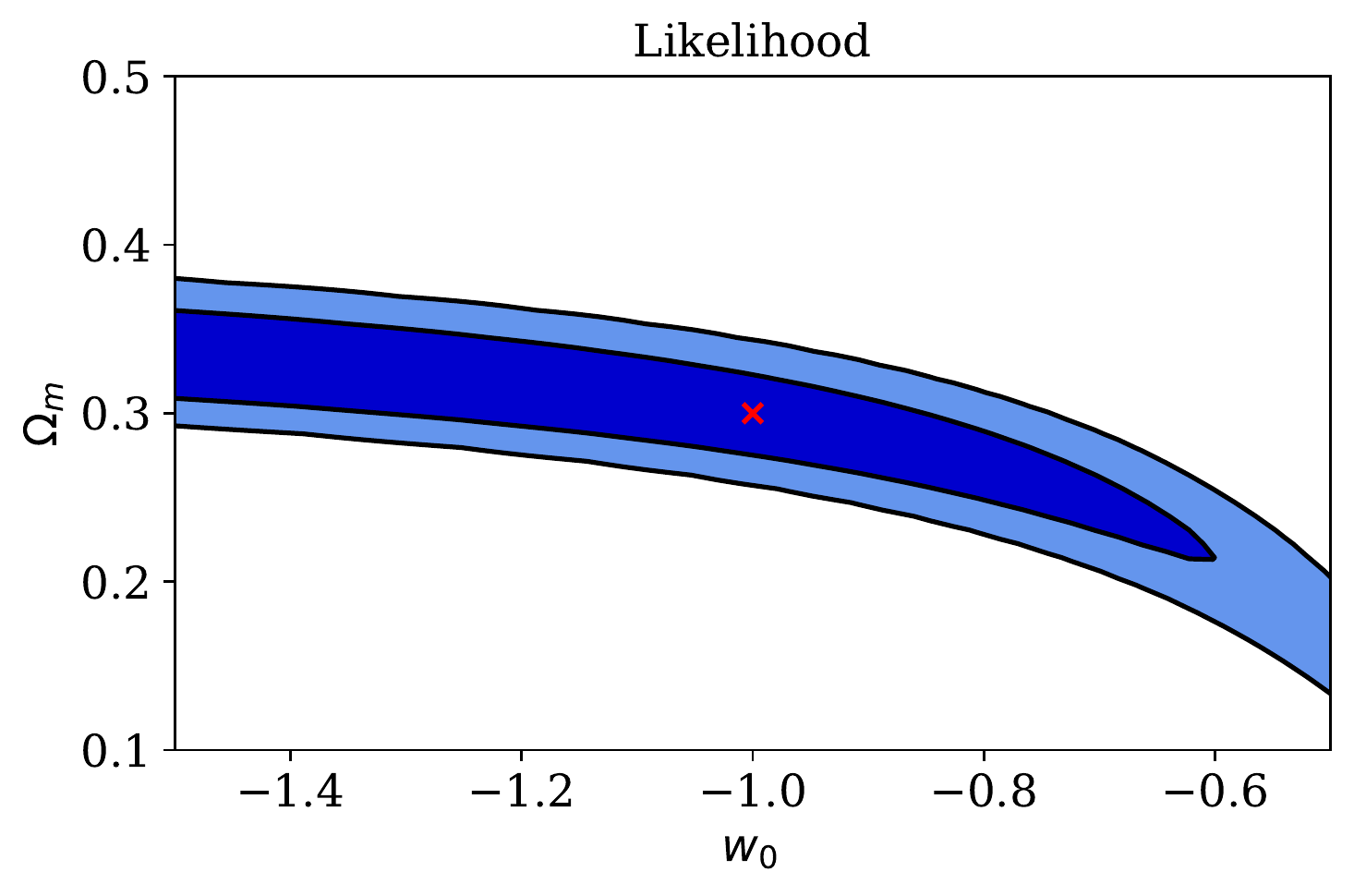}
        \end{subfigure}
    \caption[ ]
    { 2D posterior distribution in $H_0$ and $w_0$ (left) and $\Omega_M$ and $w_0$ (right) in a flat $w$CDM cosmological model ($w_a = 0$), inferred from 1,000 GW observations with $D_L^\mathrm{max} = 40$ Gpc. In each plot, the parameter not shown is fixed to its true value ($\Omega_M = 0.3$ on the left, $H_0 = 70$ km/s/Mpc on the right), and we take flat priors on the two free parameters. Contours show 68\% and 95\% credible regions.} 
    \label{fig:w2d}
\end{figure*}

\begin{figure}
    \centering
    \includegraphics[width=0.5\textwidth,keepaspectratio]{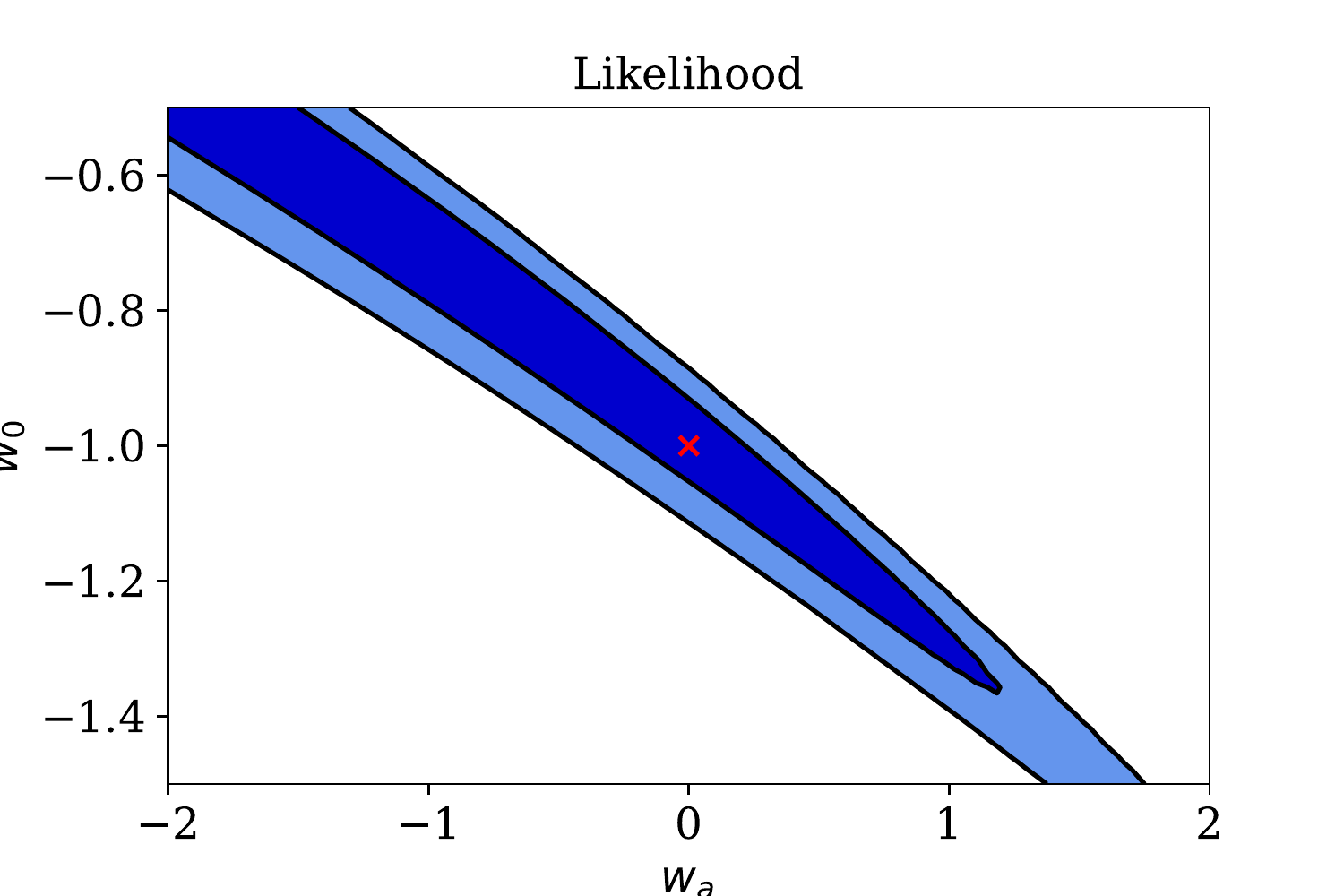}
    \caption{2D posterior distribution in $w_0$ and $w_a$ for a $w_0w_a$CDM model, inferred from 10,000 simulated luminosity distance measurements with $D_L^\mathrm{max} = 40$ Gpc. $H_0$ and $\Omega_M$ are fixed to their true values, and we adopt flat priors on $w_0$ and $w_a$. Contours show 68\% and 95\% credible regions.}
    \label{fig:w0wa}
\end{figure}

To study the ability of our proposed method to constrain cosmology, we simulate mock luminosity distance measurements according to Eq.~\ref{eq:DLobs}. We test two different detection thresholds to test the impact of the assumed $D_L^\mathrm{max}$. By default we assume that all systems with $D_L^\mathrm{obs} < 40$ Gpc are detectable, but for comparison, we also explore constraints with an observed distance limit of 20 Gpc. Given 10, 100, 1000, and 10,000 mock GW observations within the chosen distance limit, we calculate posteriors over cosmological parameters. All parameter inference is done with an input  flat $\Lambda$CDM cosmology with $H_0=70$ km s$^{-1}$ Mpc$^{-1}$, $\Omega_M=0.3$, $\Omega_{\Lambda}=0.7$. For extensions to $\Lambda$CDM, we use default values of $w_{0}=-1$, $w_{a}=0$, and $c_M=0$. We assume the merger rate evolution is known perfectly as a function of redshift according to Eq.~\ref{eq:MD}.
\subsection{$H_0$ and $\Omega_M$}
We begin by assuming a flat $\Lambda$CDM universe and calculating 2D posteriors in $H_0$ and $\Omega_M$ given our simulated distance measurements. Figure~\ref{fig:contour} shows an example posterior from 10,000 GW events, given flat priors in $\Omega_M$ and $H_0$. The 2D posterior is highly degenerate and unsurprisingly constrains $H_0$ much more strongly than $\Omega_M$. \reply{By empirically fitting the degeneracy, we find that} our method is most sensitive to the combination $H_0^{2.8}\Omega_M$, which differs from the combination $H_0^2\Omega_M$ best-measured by the CMB. This method, if used as a joint probe, can help break the degeneracy in $H_0$ and $\Omega_M$ in measurements by current or future CMB experiments.

 We estimate the expected constraints in terms of $H_0^{2.8}\Omega_M$ for different sample sizes in Fig.~\ref{fig:histograms}. We find that the convergence of the 1$\sigma$ (68\% credibility) constraint in $H_{0}^{2.81}\Omega_M$ scales with the number of events $N$ as $\frac{18\%}{N^{0.5}}$ for a distance limit of $D_L^\mathrm{max} = 40$ Gpc. For a distance limit of 20 Gpc, the expected precision is degraded to $\frac{50\%}{N^{0.5}}$. Much of the cosmology information appears to come from distances greater than 20 Gpc, as expected from Fig.~\ref{fig:dLs_cosmo}. If $H_0$ is measured at sub-percent levels from nearby BNS mergers with counterparts and the merger rate evolution is known, we expect to constrain $\Omega_M$ to the 1\% level with a couple of hundred of observations (to be expected within a few weeks of observing with 3G detectors).

\subsection{Dark Energy Parameters}

\begin{figure*}
        \centering
        \begin{subfigure}[t]{0.5\textwidth}  
            \centering 
            \includegraphics[width=\textwidth]{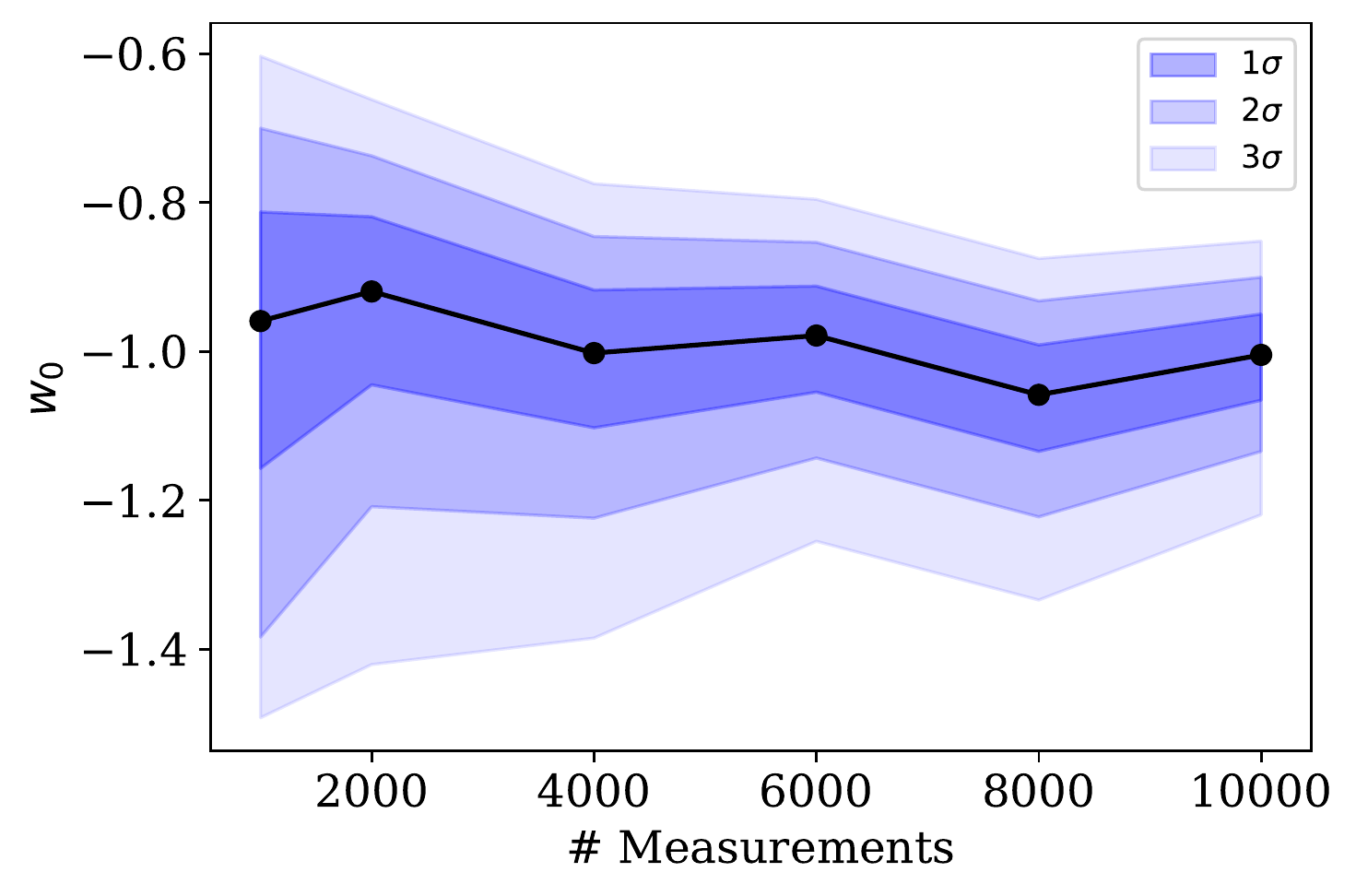}
            \caption[]%
            { Projected convergence of measurements of $w_{0}$ with fixed cosmology and $w_0=-1$.} 
    \label{fig:w0convergence}
        \end{subfigure}
        \hfill
        \begin{subfigure}[t]{0.475\textwidth}  
            \centering 
            \includegraphics[width=\textwidth]{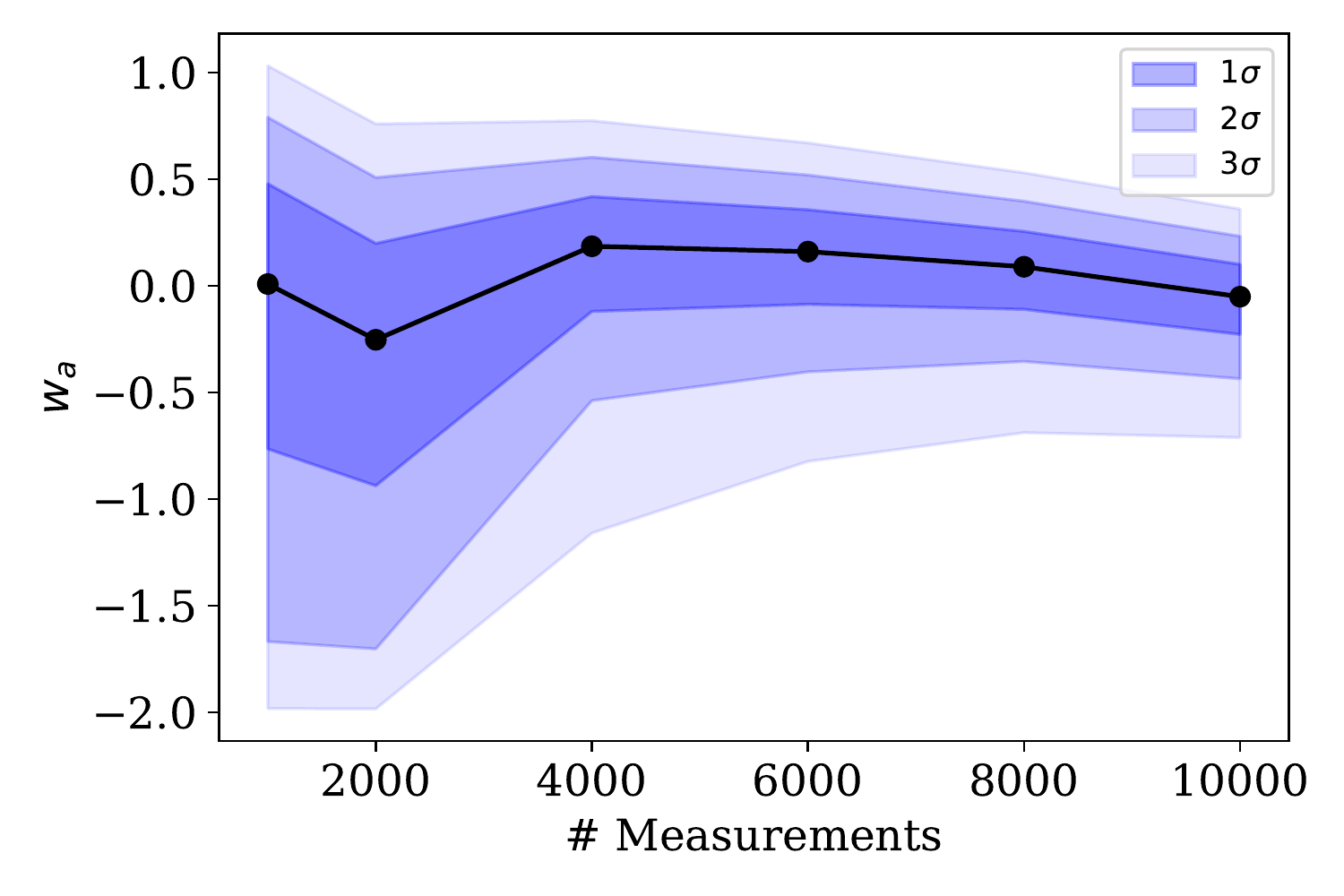}
            \caption[]%
            { Projected convergence of measurements of $w_{a}$ with fixed cosmology and $w_a=0$. Plot shows one realization of each measurement count. Constraints on $w_a$ are strongly dependent on the value of $w_a$, and so statistical fluctuations are reflected in the credible interval width} 
    \label{fig:waconvergence}
        \end{subfigure}
    \caption[ ]{Projected convergence of dark energy equation of state measurements.}
    
\end{figure*}

Next we consider extensions to flat $\Lambda$CDM and their effect on the background cosmology. We use the $w_0w_a$ parameterization of the equation of state with free parameters $w_0$ (the equation of state parameter at $z = 0$) and $w_a$ (the evolution of the equation of state with scale factor $a = \frac{1}{1+z}$). While our method is sensitive to the dark energy equation of state, the resulting constraints on the dark energy parameters are largely degenerate with measurements of $\Omega_M$ and $H_0$, \reply{which dominate the constraints}, as seen in Fig.~\ref{fig:w2d}. Nevertheless, with external cosmological priors on $H_0$ and $\Omega_M$, we can derive meaningful constraints on $w_0$ and $w_a$. Fixing $H_0=70$, $\Omega_M=0.3$, $\Omega_\Lambda=0.7$, we derive joint constraints on $w_0$ and $w_a$ in Fig.~\ref{fig:w0wa}. These two parameters are degenerate, such that a larger value of $w_0$ and a smaller $w_a$ are consistent with the input cosmology. Fixing one parameter and constraining the other, the convergence of the 1$\sigma$ constraint in $w_{0}$ scales as $\frac{500\%}{N^{0.5}}$ assuming a distance limit of 40 Gpc (see Fig. \ref{fig:w0convergence}), and also scales as $\sqrt N$ for $w_a$ in a fixed cosmology (Fig. \ref{fig:waconvergence}).
The width of the credible intervals in $w_a$ are highly dependent on the maximum prior bound considered for $w_a$, where positive $w_a$ is constrained much more strongly.
If we work with a $w$CDM model ($w_a = 0$) and adopt sub-percent prior constraints on $H_0$ and $\Omega_M$, we expect that 10,000 events can constrain the dark energy equation of state parameter $w_0$ to 5\%, comparable to, but completely independent of, the available constraints from the combination of CMB, BAO, supernovae and weak lensing data~\citep{2019PhRvD..99l3505A}.

\begin{figure*}
        \centering
            \includegraphics[width=0.75\textwidth]{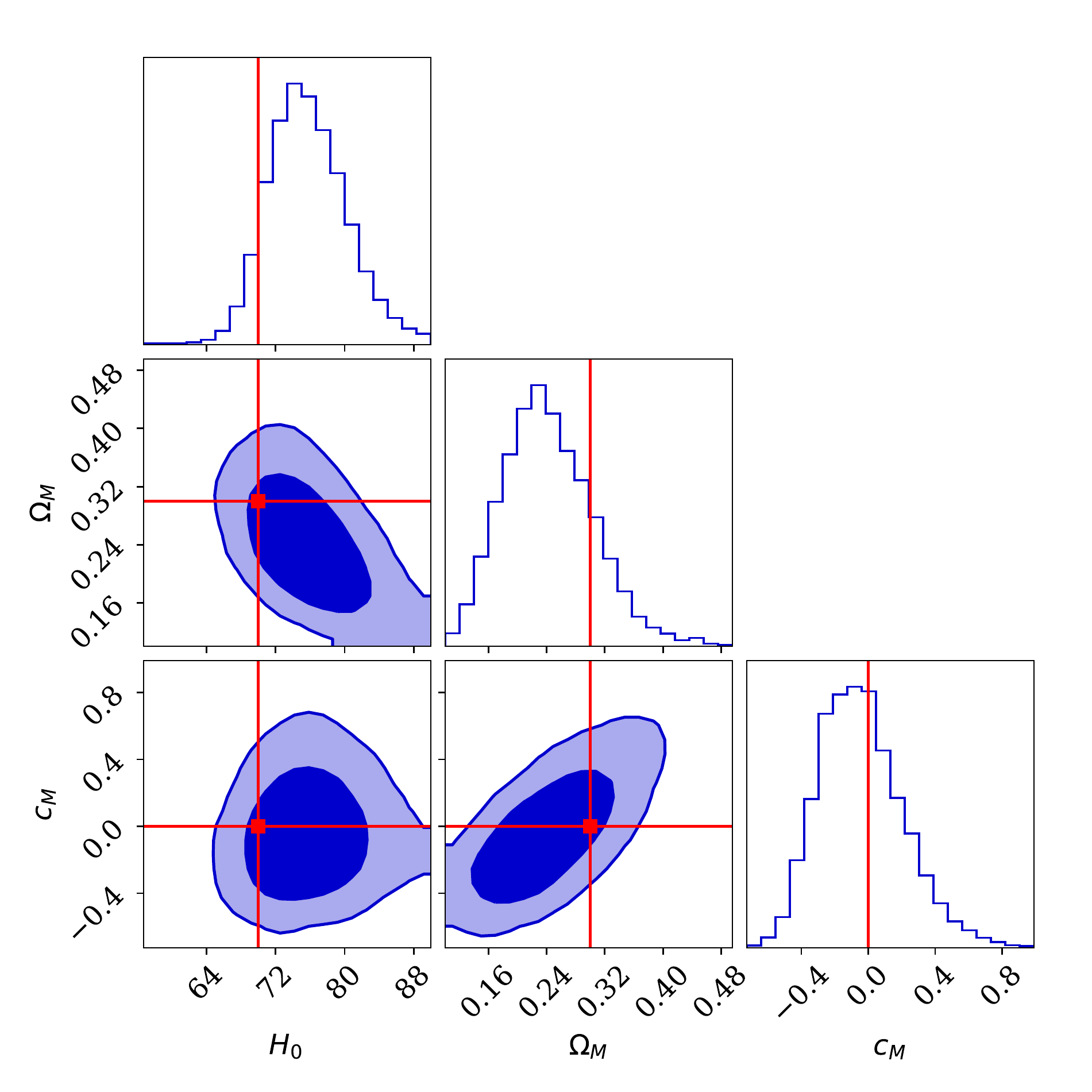}
    \caption[ ]
    {\reply{Joint posterior distribution in $H_0$, $c_M$ and $\Omega_M$ and $c_M$ in a flat $\Lambda$CDM background cosmology, inferred from 10,000 simulated distance measurements. We pick flat priors in all parameters. Contours show 68\% and 95\% credible regions.}} 
    \label{fig:cm3d}
\end{figure*}

\begin{figure*}
        \centering
        \begin{subfigure}[b]{0.475\textwidth}
            \centering
            \includegraphics[width=\textwidth]{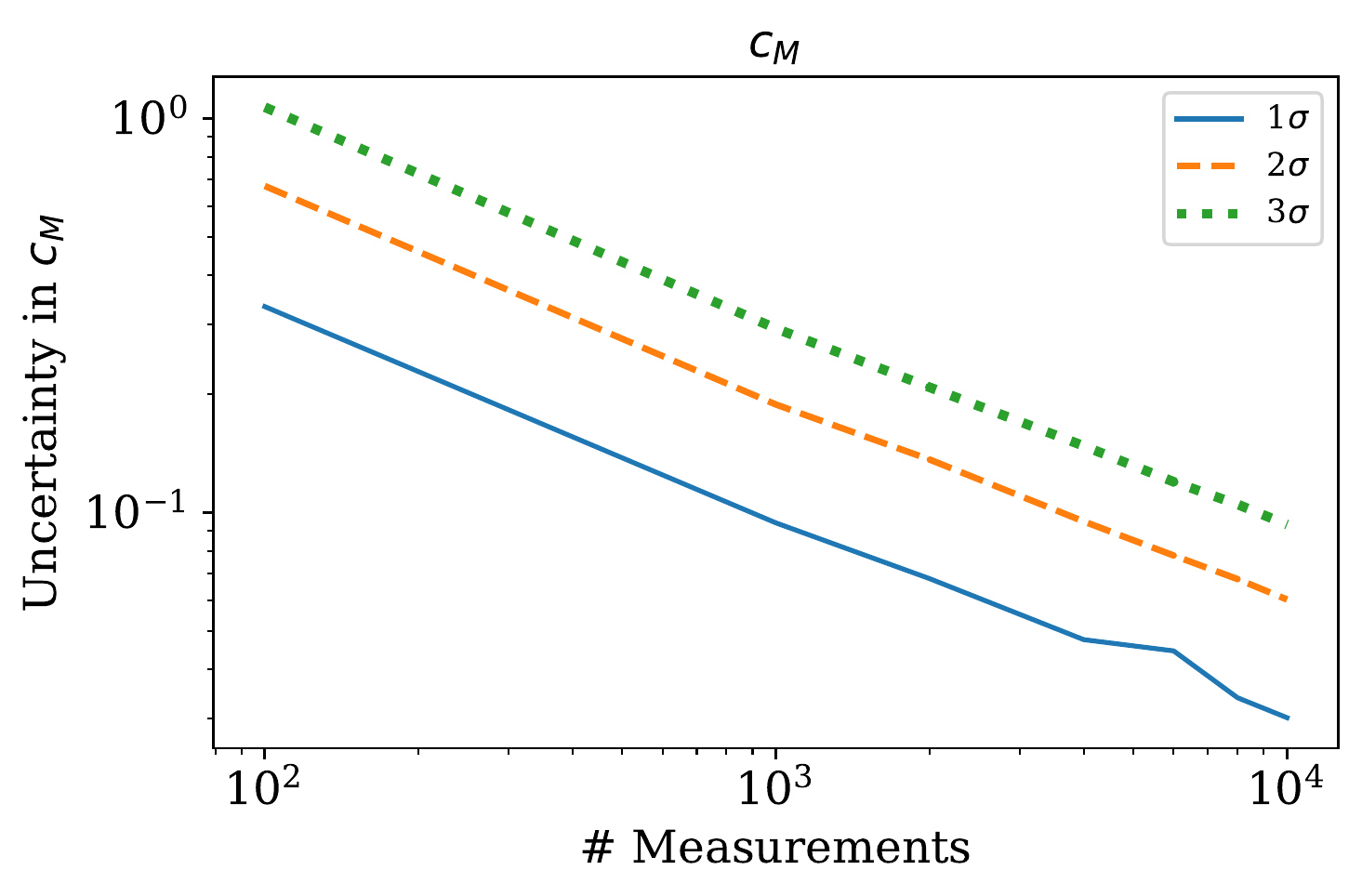}
            \caption[]%
            {{\small Uncertainty }}    
        \end{subfigure}
        \hfill
        \begin{subfigure}[b]{0.475\textwidth}  
            \centering 
            \includegraphics[width=\textwidth]{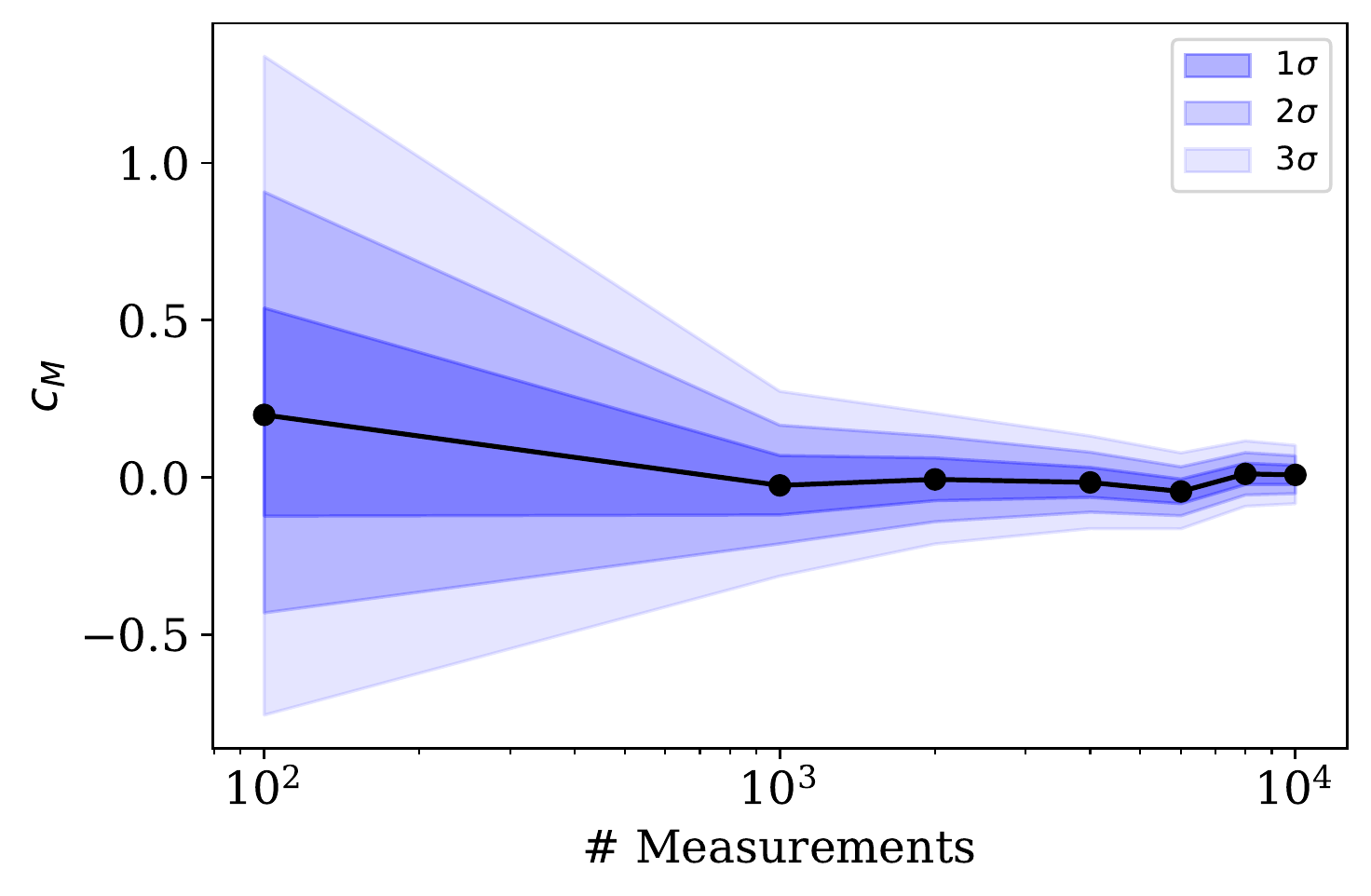}
            \caption[]%
            {{\small Credible intervals}}    
        \end{subfigure}
    \caption[ ]
    {Projected convergence of measurements of the running of the Planck mass parameter $c_{M}$ in a flat $\Lambda$CDM background cosmology. The parameters of the background cosmology are assumed to be perfectly measured.} 
    \label{fig:cmconvergence}
\end{figure*}

\begin{figure*}
\begin{subfigure}{0.3\textwidth}
\includegraphics[width=\linewidth]{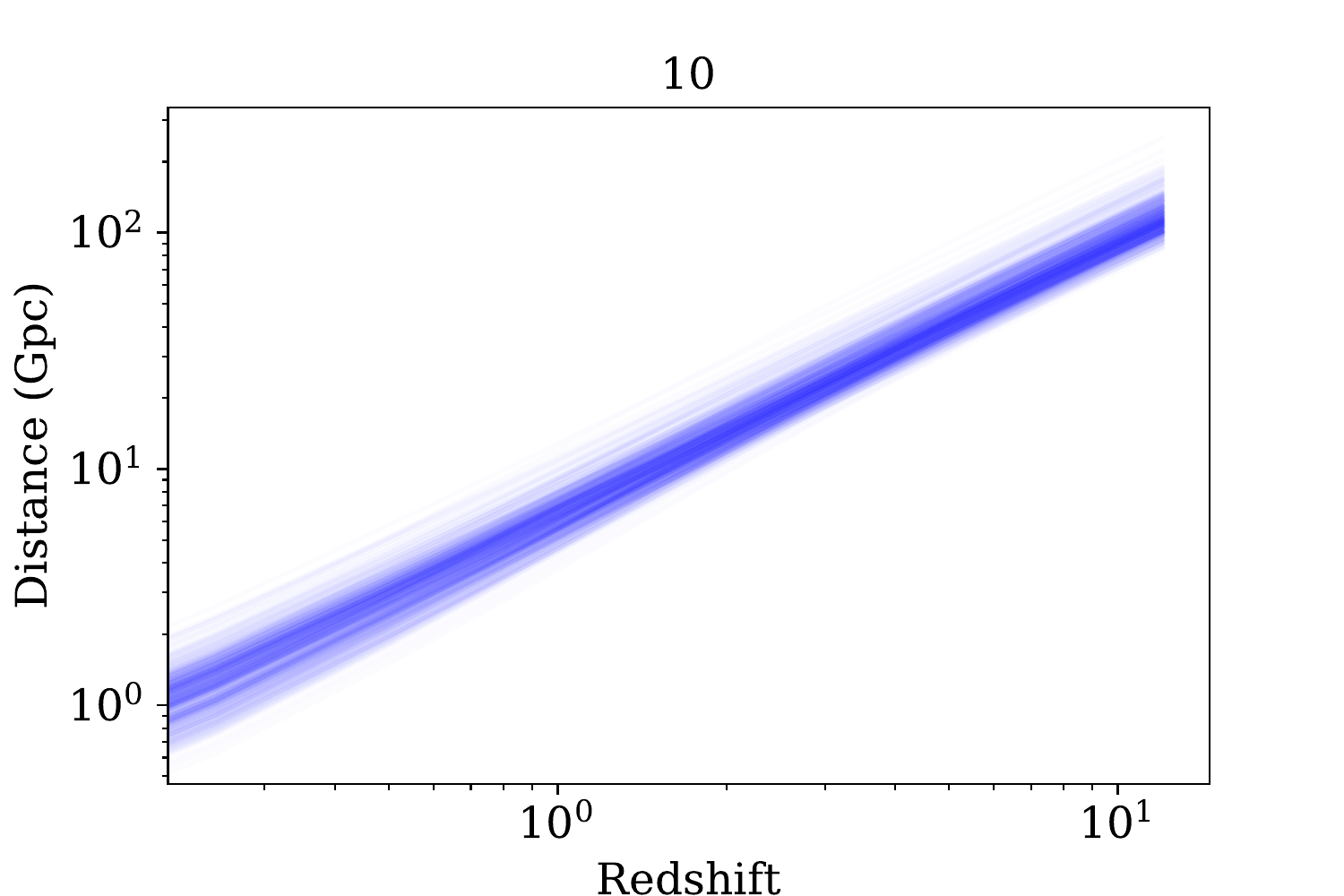}
\end{subfigure}
%\hfill
\begin{subfigure}{0.3\textwidth}
\includegraphics[width=\linewidth]{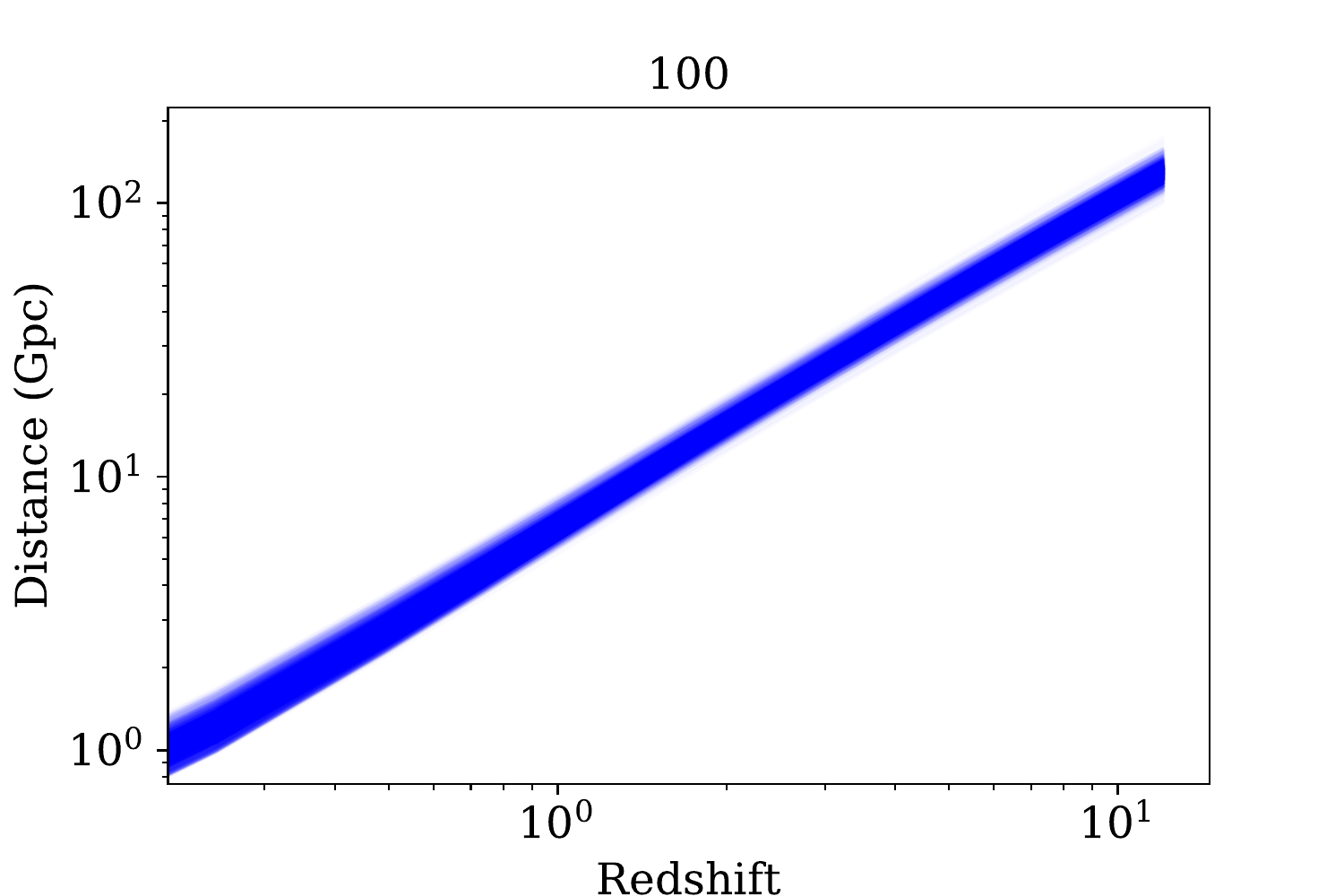}
\end{subfigure}
%\bigskip 
\begin{subfigure}{0.3\textwidth}
\includegraphics[width=\linewidth]{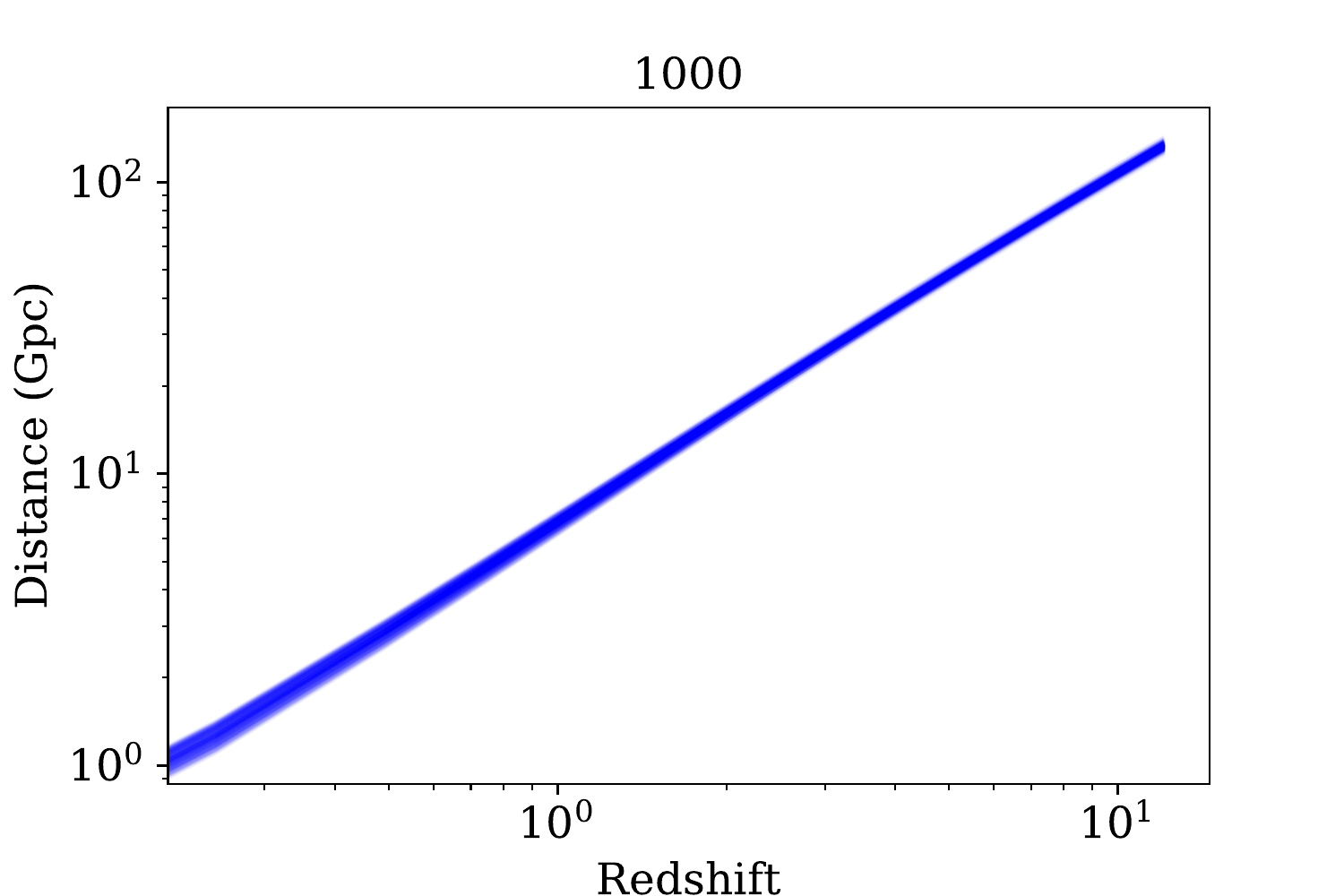}
\end{subfigure}

\caption{\label{fig:dz} Distance-redshift curves sampled from the joint posterior in $H_0$ and $\Omega_M$, calculated with a flat prior in $H_0$ and $\Omega_M$. We show the inference with 10, 100 and 1,000 events within a distance limit of 40 Gpc. The degeneracy in $H_0$ and $\Omega_M$ captures a consistent distance-redshift relation.} % Overall figure caption
\end{figure*}

\begin{figure*}
\begin{subfigure}{0.3\textwidth}
\includegraphics[width=\linewidth]{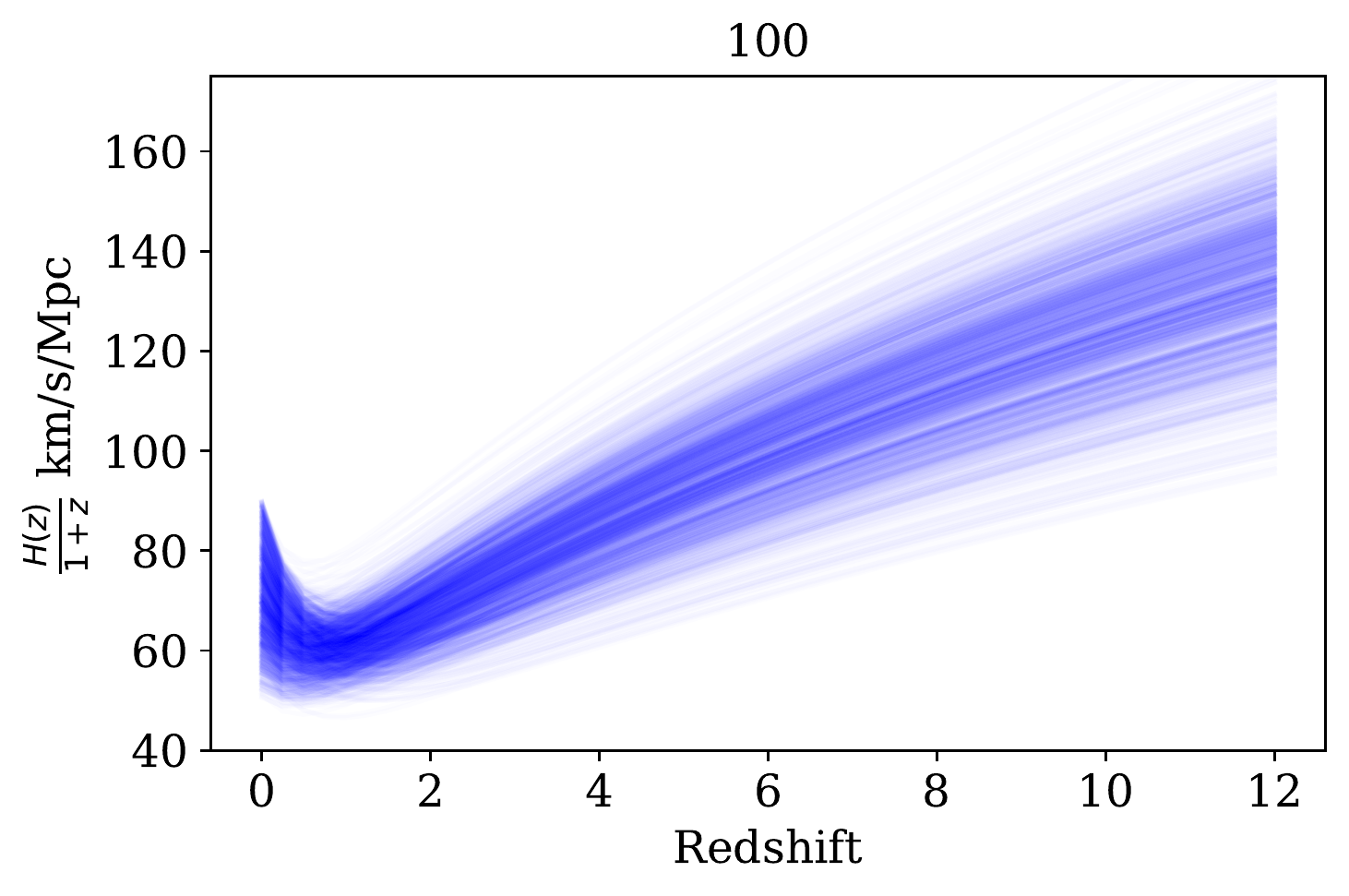}
\end{subfigure}
%\bigskip 
\begin{subfigure}{0.3\textwidth}
\includegraphics[width=\linewidth]{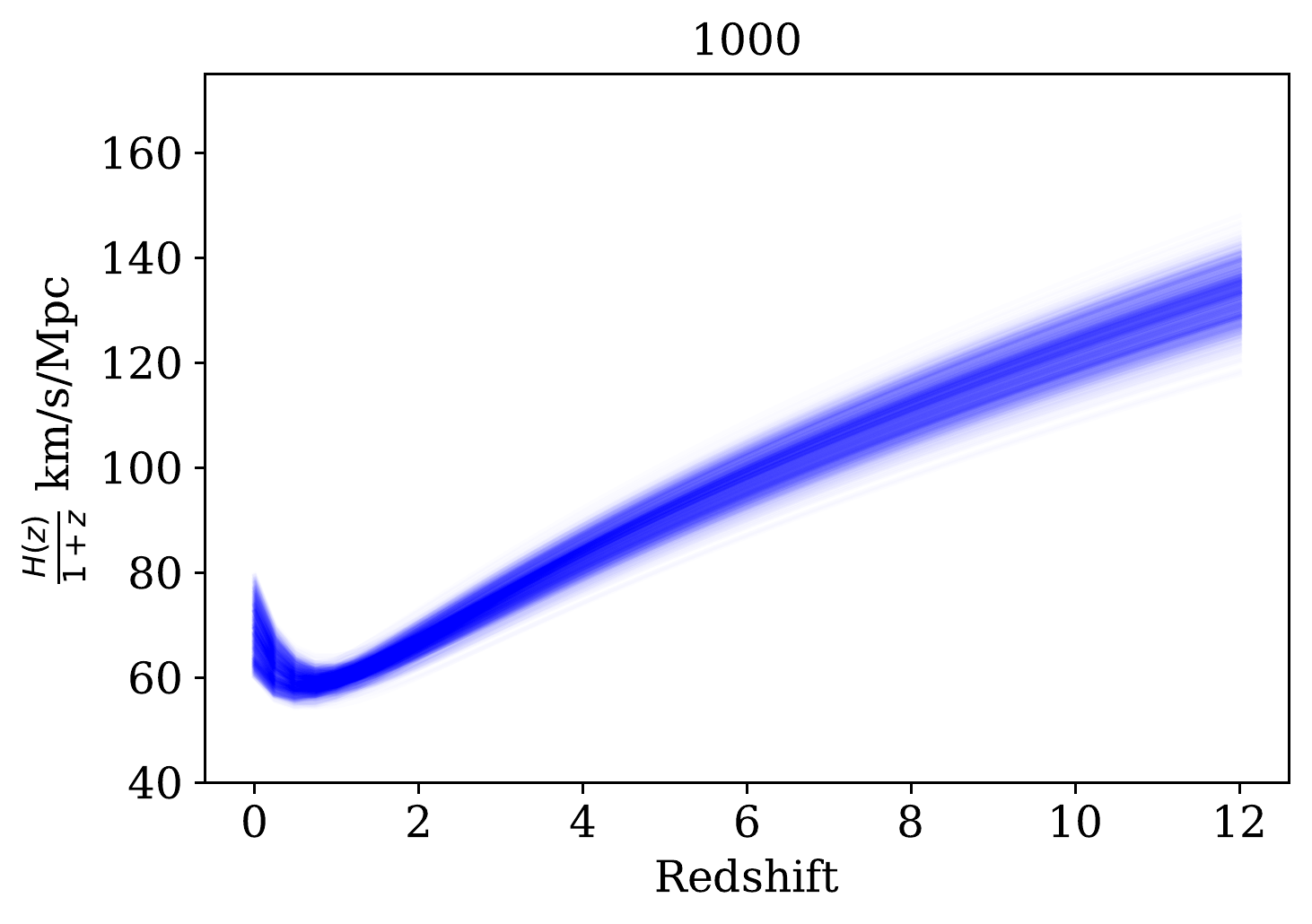}
\end{subfigure}
%\hfill % maximize the horizontal distance between the graphs
\begin{subfigure}{0.3\textwidth}
\includegraphics[width=\linewidth]{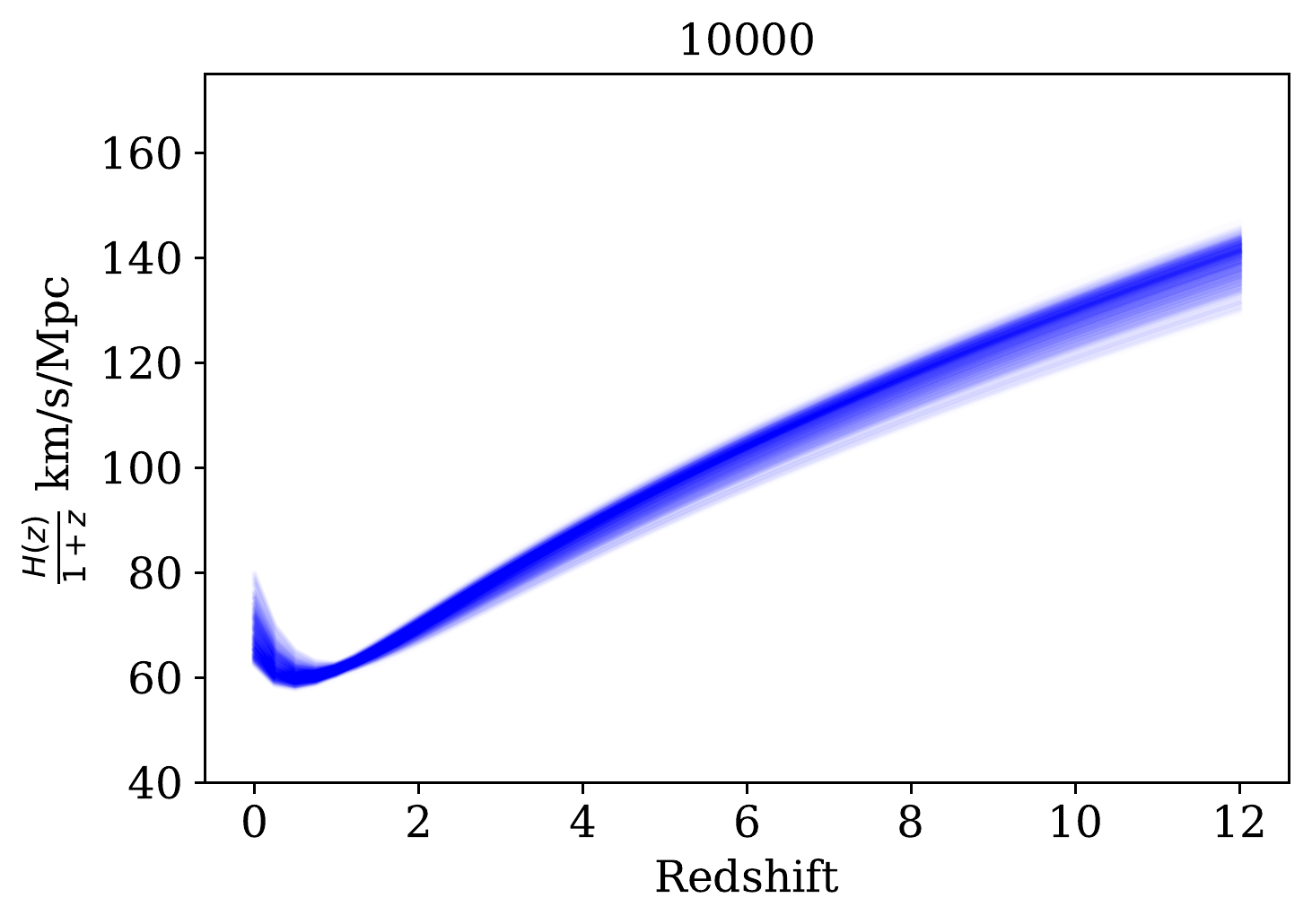}
\end{subfigure}
\caption{\label{fig:hz} $H(z)$ evolution sampled from the joint posterior in $H_0$ and $\Omega_M$ in a flat $\Lambda$CDM model, calculated with a flat prior in $H_0$ and $\Omega_M$. We show the inference with 100, 1,000 and 10,00 events within 40 Gpc. $\frac{H(z)}{1+z}$ is best constrained around $z\sim1$--$2$.} % Overall figure caption
\end{figure*}

\subsection{Modified Gravity}
Finally, we study extensions to $\Lambda$CDM through their effect on modified GW propagation. We assume the effect of the modified theory on the background expansion is minimal, so that the background expansion can be described by $\Lambda$CDM, and forecast constraints on $c_M$, the running of the Planck mass, based on the parameterization given in Eq. \ref{eq:running}. Using standard sirens, it is possible to jointly constrain $c_M$, $H_0$, and $\Omega_M$, although the joint posterior has strong degeneracies (see Fig.~\ref{fig:cm3d}). \reply{Jointly inferring $H_0$, $\Omega_M$, and $c_M$ with broad priors, the $1\sigma$ width of the marginal posterior on $c_M$ converges roughly as $\frac{60}{N^{0.5}}$.}

Fixing all other parameters, including $H_0$, the width of the 1$\sigma$ constraint in $c_M$ scales approximately as $\frac{3.4}{N^{0.5}}$, with $N$ the number of events, as shown in Fig. \ref{fig:cmconvergence}. Current cosmological measurements constrain $c_M$ to $\pm \mathcal{O}(1)$~\citep{2019PhRvD..99j3502N}, while BNS observations with counterparts in the advanced LIGO era can constrain $c_M$ to $\pm 0.5$~\citep{2019PhRvD..99h3504L}. We find that if the merger redshift distribution is known \reply{and $H_0$ and $\Omega_M$ are perfectly measured,} a hundred BNS observations within a distance limit of 40 Gpc can already surpass these projected limits. \reply{Without using external measurements on $H_0$ and $\Omega_M$, it would take $\sim 10,000$ events to surpass these limits.}
%\reply{However, as discussed in~\citet{2019PhRvD..99h3504L}, it is in practice necessary to jointly infer $H_0$, $c_M$, and $\Omega_M$, or risk a systematic bias. Jointly inferring $H_0$, $\Omega_M$, and $c_M$ with broad priors, the width of the marginal posterior on $c_M$ converges roughly as $\frac{60}{N^{0.5}}$.}
We can interpret these constraints in terms of the value of the effective Planck mass or Newton's constant at redshift $z$ compared to today~\citep{2018FrASS...5...44E,2021JCAP...02..043M}. For $c_M = 0$ \reply{and fixed $H_0$ and $\Omega_M$}, the 1$\sigma$ measurement in $c_M$ from 10,000 GW events translates to an effective Planck mass of $2.172 \pm{ 0.017} \times 10^{-8}$ kg, or an effective Newton's constant of $6.70 \pm{0.11} \times 10^{-11} \frac{\mathrm{N} \cdot \mathrm{m}^2}{\mathrm{kg}^2}$ at $z=2$. \reply{Additionally, we can repeat the analysis using the modified GW propagation model proposed by~\citet{2018PhRvD..98b3510B}, parameterized in terms of $\Xi_0$ and $n$. As an example, we fix $n=1.91$, as predicted by the RT nonlocal gravity model~\citep{Maggiore_2014,2021arXiv210112660F}. With all other cosmological parameters fixed, a simulated 10,000 events yields a measurement $\Xi_0 = 1.002 \pm{0.009}$ ($\Xi_0=1$ for GR).} These measurements at $z \sim 2$ could complement observations by the Laser Interferometer Space Antenna (LISA), which will probe modified GW propagation out to even higher redshifts ($z \lesssim 10$) by observing GWs from supermassive BBH mergers with possible EM counterparts~\citep{2021JCAP...01..068B}.

\subsection{Discussion}
Comparing a catalog of GW luminosity distances against a known redshift distribution is ultimately sensitive to the underlying distance-redshift relation, as also pointed out by \citet{2019JCAP...04..033D}. For the flat $\Lambda$CDM and $w$CDM models also considered by \citet{2019JCAP...04..033D}, we find similar results for the expected constraints on $H_0$, $\Omega_M$ and $w$ with 10,000 events (compare their Fig. 2 with our Fig.~\ref{fig:contour}, for example).
Regardless of the assumed cosmological model, which provides a parameterization for the distance-redshift relation, we can examine our parameter measurements from the previous subsections in terms of constraints on the luminosity distance-redshift relation or \reply{the expansion rate} $\frac{H(z)}{1 + z}$.
Fig. \ref{fig:dz} shows posterior draws from the distance-redshift relation inferred in a flat $\Lambda$CDM model with flat priors on $H_0$ and $\Omega_M$. Draws of $H_0$ and $\Omega_M$ within our posterior are such that the luminosity distance is the same for a given redshift, and so $H_0$ has a dominant effect. 

Drawing $H_0$ and $\Omega_M$ from the joint posterior, we also look at the expected constraints on the $H(z)$ evolution as a function of redshift, as in Figure \ref{fig:hz}. The spread in $\frac{H(z)}{1+z}$ is smallest at redshifts $z \gtrsim 1$. In a $w_0w_a$CDM model, the joint posterior in $w_0$ and $w_a$ with fixed $H_0$ and $\Omega_M$ yields the lowest spread in $\frac{H(z)}{1+z}$ (at a non-zero redshift) at around $z=2$. This is consistent with our expectations that most of the cosmological information comes from knowledge of the redshift at which the merger rate peaks.

The forecasts described in this section depend on the true redshift distribution of GW sources, and how well it can be measured. \reply{Motivated by recent measurements that favor short delay times for BNS mergers~\citep{2014MNRAS.442.2342D,2016A&A...594A..84G,2019MNRAS.486.2896S},} we have assumed that the BNS rate density peaks around $z = 2$ like the SFR. \reply{A recent analysis of {\it Fermi} and {\it Swift} short GRBs finds that their rate density peaks between $z \sim 1.5$--$2$~\citep{2016A&A...594A..84G}. While current constraints on the BNS merger rate evolution are broad, as discussed in Section~\ref{sec:intro}, we expect the measurements to improve significantly over the next decade with upcoming observations of GRBs, kilonovae, and BNS host galaxies.} Because we expect to best constrain the cosmological expansion rate near the peak redshift, if it turns out that time delays are long and the peak is at $z < 2$, our projected constraints will differ. Crucially, if the wrong redshift evolution is assumed, the resulting cosmological inference will be biased, as explicitly demonstrated in \citet{2019JCAP...04..033D}. \reply{We therefore expect that the redshift evolution will be inferred jointly with the cosmological parameters, so that its uncertainty can be marginalized over.}

Additionally, most of our forecasts have assumed that all BNS mergers within an observed distance of 40 Gpc can be detected, and we have shown that we expect worse constraints, typically by a factor of a few, if the observed distance limit is lowered to 20 Gpc. The sensitivities of the proposed next-generation GW detectors are not yet finalized, and we expect this to affect the projections here, modifying the number of events needed to reach the desired accuracy in the cosmological parameters. 
Finally, we have considered the case in which the merger rate density $\mathcal{R}(z)$ is directly measured, rather than $p(z)$. Because of the cosmological dependence of the comoving volume element, if $\mathcal{R}(z)$ is perfectly known, there is cosmological information in $p(z)$. This effect is subdominant to the distance-redshift relation probed by the GW luminosity-distance relation, and only affects $\Omega_M$ and to a lesser extent $w_0$ and $w_a$. We expect our results to differ slightly in the case that $p(z)$ is more directly available. 

Standard sirens are an independent probe to address the tension in $H_0$ measurements between so-called `early-universe' and `late-universe' estimates. While with a flat prior, $H_0$ and $\Omega_M$ are strongly degenerate, a precise measurement of $H_0$ is possible with our method using an outside prior on $\Omega_M$, such as from measurements of the CMB, galaxy clustering, or weak lensing. Given that the joint posterior in $H_0$ and $\Omega_M$ is captured by $H_0^{2.8}\Omega_M$, when used with experiments sensitive to a different combination of $H_0$ and $\Omega_M$, our method can help break this degeneracy. Standard sirens are also uniquely poised to probe the nature of dark energy, not only through its effect on the background expansion parameterized by the dark energy equation of state $w$, but primarily on its effect on GW propagation, parameterized by $c_M$ here. To constrain the dark energy parameters $w_a$ and $w_0$, or the running of the Planck mass in modified gravity $c_M$, outside priors on both $H_0$ and $\Omega_M$ are necessary to reveal the sub-dominant effects on the GW distance distribution.

\section{Conclusion}
\label{sec:conclusion}
GW standard sirens can independently test the $\Lambda$CDM cosmological model and provide insight into the mysterious dark sector, namely dark matter and dark energy. 
The next generation of GW detectors, the proposed Einstein Telescope and Cosmic Explorer, would revolutionize standard siren science by observing the GW universe out to tens of Gpc. The challenge for GW cosmology will be to measure the redshifts of these mergers, especially considering the difficulties of identifying EM counterparts and potential host galaxies at $z \gg 1$. 

Previous work~\citep{2019JCAP...04..033D} showed that, in the absence of targeted EM followup campaigns or complete galaxy catalogs, prior knowledge of the \emph{distribution} of merger redshifts can be compared against GW luminosity distances to infer cosmological parameters. In this work we argue that we can leverage external measurements of the evolution of the BNS merger rate, which, in particular, is expected to peak at some redshift. This provides a redshift feature which can be used in a standard siren analysis to constrain cosmology and modified gravity. As a demonstration of this method, we used a simple toy model in which the evolution of the BNS merger rate as a function of redshift is known perfectly, and studied how the observed GW luminosity distance distribution alone can measure parameters of the $w_0w_a$CDM model and the running of the Planck mass. This allows us to isolate the available information in a catalog of GW distances, compared to the additional information that enters from the mass distribution. 

In reality, we expect this method to be used jointly with fits to the mass distribution and/or available galaxy information. The information from the mass distribution will likely dominate the inference if there is a sharp, redshift-independent mass feature like a NS-BH mass gap at low masses~\citep{2012PhRvD..85b3535T} or a pair-instability mass gap at high masses~\citep{2019ApJ...883L..42F}. Because the GW luminosity distance distribution inherently carries information about cosmology, even if it is not used as the primary observable to measure cosmology, it must be taken into account in all standard siren analyses at high redshifts to avoid biasing the cosmological constraints~\citep{Mortlock:2018azx,2019arXiv190806060T,2021arXiv210112660F,2021arXiv210314663M}.

We have focused on the next generation of detectors in our analysis because they will likely observe GW mergers past cosmic noon, or the peak redshift of the merger rate, providing a clear feature whose feature can be measured in both redshift and distance space. Similar analyses can in principle be carried out on existing GW catalogs; in combination with measurements of the stochastic GW background, current GW observatories will constrain the peak of the BBH merger rate distribution~\citep{2020ApJ...896L..32C}. However, currently the distance distribution is only meaningfully constrained for high-mass BBH mergers, while the corresponding redshift distribution is not well-constrained from EM observations. Existing BBH observations can only constrain large deviations from GR; for example, GW leakage in large extra dimensions~\citep{2016CQGra..33p5004C,2018ApJ...863L..41F}.

\acknowledgments
We thank Tessa Baker, Giuseppe Congedo, Xuheng Ding, Bryan Gillis and Simone Mastrogiovanni for their helpful comments on the manuscript. M.~F. is supported by NASA through NASA Hubble Fellowship grant HST-HF2-51455.001-A awarded by the Space Telescope Science Institute.

\bibliographystyle{apsrev}
\bibliography{references}

\end{document}